\documentclass[twocolumn,superscriptaddress,amsmath,amssymb,aps,prx]{revtex4-2}  
\usepackage{listings}
\usepackage{graphicx}
\usepackage{dcolumn}
\usepackage{amsmath}
\usepackage{bm}
\usepackage{ulem}
\usepackage{natbib}
\usepackage[colorlinks,citecolor=red,linkcolor=blue,urlcolor=blue]{hyperref}
\usepackage{capt-of}
\usepackage{hhline}
\usepackage{float}
\usepackage{calrsfs}
	\DeclareMathAlphabet{\pazocal}{OMS}{zplm}{m}{n}
\usepackage{physics}
\usepackage{appendix}
\usepackage[export]{adjustbox}

\usepackage[dvipsnames]{xcolor}
\definecolor{pink}{rgb}{0.858, 0.188, 0.478}

\usepackage{lineno,blindtext}
\usepackage{orcidlink}

\begin{document}

\title{Quantum Skyrmion Liquid}

\author{Dhiman Bhowmick\,\orcidlink{0000-0001-7057-1608}}
\affiliation{School of Physical and Mathematical Sciences, Nanyang Technological University, Singapore}
\affiliation{Department of Physics, National University of Singapore, Singapore 117542}

\author{Andreas Haller\,\orcidlink{0000-0003-0420-2446}}
\affiliation{Department of Physics and Materials Science, University of Luxembourg, 1511 Luxembourg, Luxembourg}

\author{Deepak S. Kathyat\,\orcidlink{0000-0002-5685-0085}}
\affiliation{School of Physical and Mathematical Sciences, Nanyang Technological University, Singapore}

\author{Thomas L. Schmidt\,\orcidlink{0000-0002-1473-3913}}
\affiliation{Department of Physics and Materials Science, University of Luxembourg, 1511 Luxembourg, Luxembourg}

\author{Pinaki Sengupta\,\orcidlink{0000-0003-3312-2760}}
\affiliation{School of Physical and Mathematical Sciences, Nanyang Technological University, Singapore}

\date{\today}

\begin{abstract}
Magnetic skyrmions are topological objects that have been observed in helimagnets. Mostly treated as classical magnetic textures, they are studied mainly for spintronics applications due to their topological stability.
Recent experimental observations of skyrmions with length scales comparable to the interatomic lattice spacing have led to a stronger focus on their quantum nature.
Despite experimental and theoretical evidence showing nanoscale quantum skyrmions, it remains unclear what physical phenomena differentiate a classical from a quantum skyrmion.
We present numerical evidence for the existence of a quantum skyrmion liquid (SkL) phase in quasi-one-dimensional lattices which has no classical counterpart.
The transition from a conventional quantum skyrmion crystal (SkX) to a field-polarized phase (FP) is found to be of second order while the analogous classical transition near zero temperature is first-order due to a missing SkL phase.
As an indicator of the quantum mechanical origin of the SkL phase, we find concentrated entanglement (indicated by the concurrence) around the skyrmion center, which we attribute to the uncertainty in the skyrmion position resulting from the non-commutativity of the skyrmion coordinate operators. The latter also gives rise to a nontrivial kinetic energy in the presence of an atomic lattice.
The SkL phase emerges when the kinetic energy dominates over the skyrmion-skyrmion interaction energy. It is tied to the breaking of discrete translational invariance of the skyrmion crystal and occurs when the skyrmion radius is comparable with the size of the magnetic unit cell.
In contrast to the long-range order present in the SkX phase, spin-spin correlations in the SkL phase exponentially decay with distance, indicating the fluid-like behavior of uncorrelated skyrmions.
The emergence of kinetic energy-induced quantum SkL phase serves as a strong indication of the possible Bose-Einstein condensation of skyrmions in higher-dimensional systems.
Our findings are effectively explained by microscopic theories like collective coordinate formalism and trial wave functions, effectively enhancing our understanding of the numerical findings. 
\end{abstract}

\maketitle

\section{\label{Section::Introduction}Introduction}

Magnetic skyrmions are vortex-like spin textures with a real-space topological charge that have attracted widespread interest ever since their discovery more than a decade ago\,\cite{SkyrmionDiscovery1, SkyrmionDiscovery2}.
The topological protection against local defects makes them suitable for storing, transporting, and processing information in spintronic devices~\cite{SkyrmionApplication1, SkyrmionApplication2, SkyrmionApplication3, SkyrmionApplication4, SkyrmionApplication5, SkyrmionApplication6, SkyrmionApplication7, SkyrmionApplication8, SkyrmionApplication9}.
Experimentally, skyrmions are mostly realized in non-centrosymmetric magnets, including bulk systems such as MnSi and FeGe~\cite{FeGeMnSi1,FeGeMnSi2,FeGeMnSi3,FeGeMnSi4,FeGeMnSi5} but also layered heterostructures such as Fe/Co/Ir~\cite{HeteroStructure1, HeteroStructure2, HeteroStructure3}.
Generically, skyrmions emerge in these systems from the interplay between the symmetric Heisenberg exchange interaction and the antisymmetric Dzyaloshinskii-Moriya interaction and are stable over a wide range of temperatures and applied magnetic fields.

Made up of quantum spins, skyrmions are fundamentally quantum mechanical in nature. However, they are primarily studied from a classical perspective because classical micromagnetics provides a good representation of the static and dynamic properties of many skyrmion systems.
This is because typical thermally activated skyrmions are large and span tens to hundreds of spins.
A skyrmion can be described in the collective coordinate formalism by the following classical Lagrangian~\cite{Tserkovnyak, LeonBalents},
\begin{equation}
\pazocal{L}_{\text{eff}}=\frac{2\pi S^\prime\pazocal{N}}{A_c}\hat{z}\cdot\left(\dot{\boldsymbol{X}}\times \boldsymbol{X}\right)-V(\boldsymbol{X}),
\label{Eq.1}
\end{equation}
where $S^\prime=\hbar S$ is the spin, $\pazocal{N}$ is the skyrmion number, $A_c=a^2$ is the area of unit-cell and $\boldsymbol{X}=(X_0,\,Y_0)$ is the position of the skyrmion center.
Moreover, $V(\boldsymbol{X})$ is the external potential experienced by a skyrmion.
The first term can be rewritten as $\dot{\boldsymbol{X}}\cdot\pazocal{A}$ and thus denotes a coupling of the skyrmion velocity to an artificial magnetic vector potential $\pazocal{A}=(2\pi S^\prime \pazocal{N}/A_c)(Y_0\hat{e}_x-X_0\hat{e}_y)$.
In this approximation, a skyrmion is interpreted as a classical particle that lacks kinetic energy and experiences an artificial magnetic flux arising from its non-trivial spin texture.

Recently, atomic scale skyrmions have been realized in several materials~\cite{AtomicSkyrmion1, AtomicSkyrmion2, AtomisticSkyrmion3, AtomisticSkyrmion4, AtomisticSkyrmion5, Nyayabanta}.
It is expected that quantum effects lead to significant corrections when the size of the skyrmion becomes comparable to the lattice constant $a$ of the underlying atomic lattice and the spin $S$ of the skyrmion is small.
Theoretically, several studies have investigated the presence of quantum skyrmions in spin exchange models~\cite{AndreasHaller, AchimRosch, Model_Joshi2023, Model_Siegl2022, haller2024quantum}, proposed detection schemes to probe them~\cite{Measurement_Sotnikov2021, Measurement_Sotnikov2023, Measurement_Salvati2023_unpublished}, studied their stability and dynamics~\cite{Stability_Vlasov2020,Measurement_Salvati2023_unpublished,Dynamics_Psaroudaki2017, Dynamics_Vijayan2023,SoOn_Diaz2021}, their potential use as qubits~\cite{Qbit_Psaroudaki2021, Qbit_Xia2023} or as mobile magnetic impurity platforms to perform topological quantum computation with Majorana fermions~\cite{SoOn_Diaz2021b, SoOn_Nothhelfer2022}.

\begin{figure}[tb]
\includegraphics[width=0.45\textwidth]{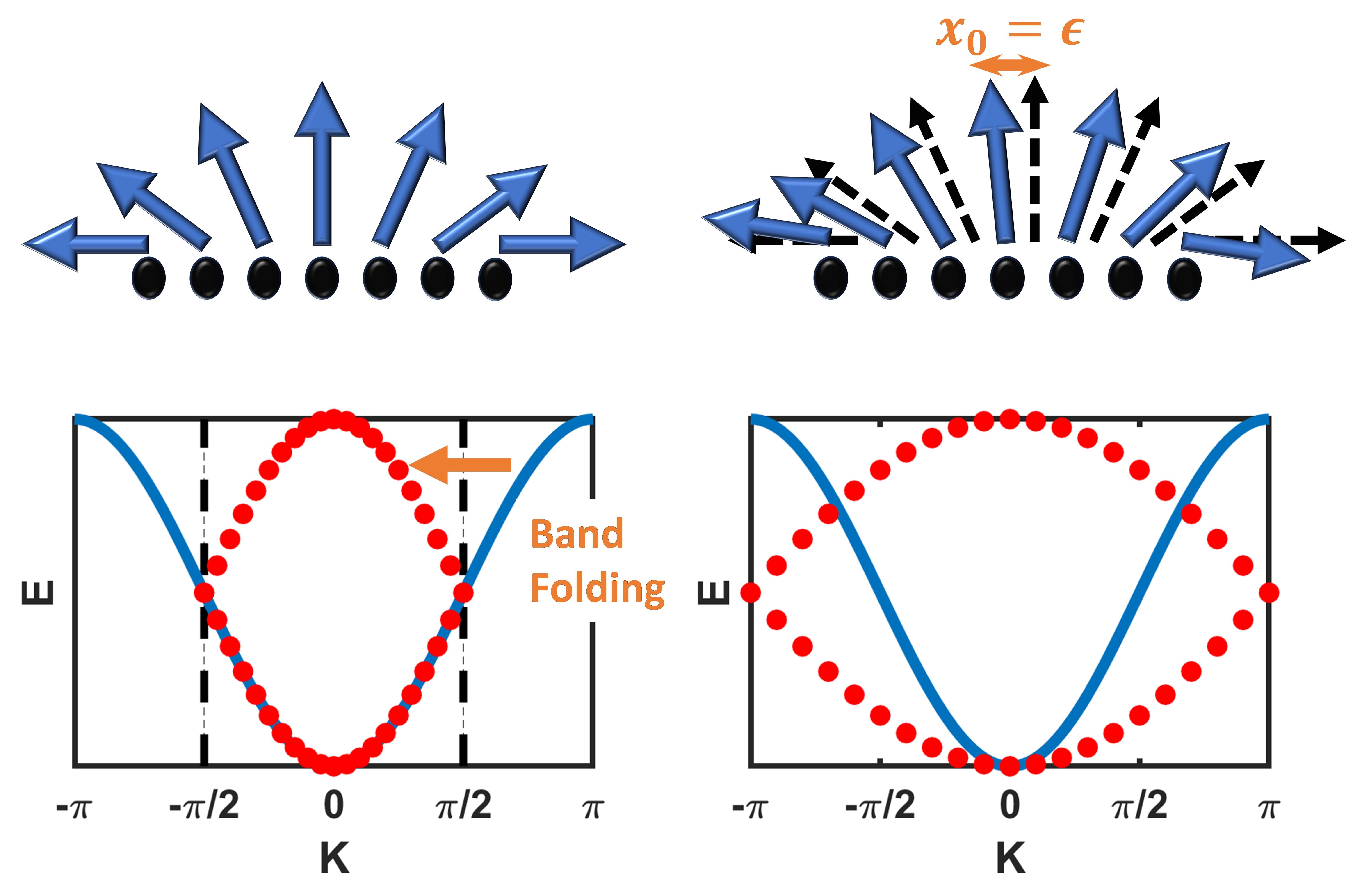} 
\caption{Schematic representation of the translation of a skyrmion over a discrete crystal (top) and skyrmion band structure (bottom). \textit{Top:} The left figure shows the skyrmion center overlapping with a crystal site. The right figure shows a skyrmion with the center shifted by $\epsilon$ towards the right\,(dotted black lines). The energy of the two configurations is different, resulting in a periodic potential of a skyrmion. For a large skyrmion with radius $R\gg a$ (classical limit), the background lattice acts like a continuum, resulting in a negligible lattice potential. \textit{Bottom:} The blue line shows a schematic band structure of a charged particle (left) or skyrmion (right) when the magnetic Brillouin zone is the same as the lattice Brillouin zone. On the other hand, the red dots denote the band structure of a charged particle (left) or skyrmion (right) when the magnetic Brillouin zone is different from the lattice Brillouin zone. The mismatch between the lattice and magnetic Brillouin zone results in band-folding at the magnetic Brillouin zone and at the lattice Brillouin zone for the charged particle and the skyrmion, respectively. In the case of skyrmions, the number of bands within a certain energy interval is $N=2S\pazocal{N}$, indicating infinitesimal bandwidth in the classical spin limit $S\rightarrow \infty$ resulting in a zero kinetic energy of classical skyrmions.}
\label{fig::Schematic}
\end{figure}

In this work, we focus on a different aspect concerning the interplay between quantum skyrmions and the background lattice.
In the case of a large skyrmion with radius $R\gg a$, moving the spins across the lattice costs only a small energy compared to the total energy of the system, so discretization effects can be neglected.
Instead, when the radius of the skyrmion approaches $R\approx a$, the energetic cost of such translations cannot be neglected (see top of Fig.~\ref{fig::Schematic}) and results in a non-negligible periodic potential $V(\boldsymbol{X})$ for the collective skyrmion coordinates.
Using the Lagrangian~\eqref{Eq.1}, it can be shown that for a classical skyrmion such a potential leads to a finite velocity perpendicular to the potential profile,
\begin{equation}
    4\pi S\pazocal{N}\dot{\boldsymbol{X}}\times \hat{z} =-\frac{\partial V}{\partial \boldsymbol{X}},
\end{equation}
which is known as the Magnus effect.
For quantum skyrmions, the potential profile along a particular direction can be reinterpreted as a translation operator in this direction, which is the quantum analog of the classical Magnus effect. 
Thus, as a consequence of the kinetic energy generated by the translation operators in a periodic lattice potential, quantum skyrmions feature a dispersive band structure.
The lattice potential and the corresponding skyrmion band structure become more dispersive for a smaller radius of the skyrmion. 
Rina \textit{et al.} have proposed that a dispersive skyrmion band structure will eventually lead to Bose-Einstein condensation (BEC) of skyrmions in two-dimension~\cite{LeonBalents}.

In addition, trapping a skyrmion in a certain direction will lead to a gain in kinetic energy in the perpendicular direction, which eventually indicates the non-commutativity of skyrmion position operators and leads to uncertainty in the skyrmion position.
This non-commutative relation can be derived by quantizing the Poisson bracket derived from the Lagrangian~\eqref{Eq.1},
\begin{equation}
    \left\lbrace X_0, Y_0\right\rbrace=\frac{A_c}{4\pi S^\prime\pazocal{N}}
    \implies 
    \left[ X_0, Y_0\right]=\frac{iA_c}{4\pi S\pazocal{N}}.
\end{equation}
This relation is equivalent to the commutation relation of the guiding center of a charged particle in a magnetic flux and is associated with a magnetic Brillouin zone which is $N=2S\pazocal{N}$ times larger than the size of the lattice Brillouin zone.
As a result, the skyrmion spectrum consists of $N$ bands within a finite energy range (see bottom of Fig.~\ref{fig::Schematic}).
For a fixed skyrmion radius but a higher spin $S$, the bandwidth of each band decreases and eventually vanishes for $S\rightarrow\infty$, which coincides with the classical limit. 
Thus, in the classical spin limit, due to a lack of kinetic energy quantum mechanical phenomena such as a BEC of skyrmions also cease to exist.

Such quantum effects of skyrmions were studied by Hector \textit{et al.}~\cite{Tserkovnyak} and Rina \textit{et al.}~\cite{LeonBalents} in the collective coordinate formalism.
Nevertheless, this formalism is limited to a single-particle description of skyrmions and neglects their internal (quantum) degrees of freedom as well as the skyrmion-skyrmion interaction.
In our paper, we investigate the ground state phases of interacting quantum skyrmions in a quasi-one-dimensional (1D) geometry using density matrix renormalization group (DMRG) simulations~\cite{DMRG1, DMRG2, DMRG3, DMRG4, DMRG5}.
Our results point to the emergence of a quantum skyrmion liquid (SkL) phase over a finite range of applied magnetic fields between the quantum skyrmion crystal (SkX) and field-polarized (FP) phases.
We characterize the SkL phase extensively by calculating the spatial profiles of the magnetization, correlation functions and the entanglement (concurrence) distributions.
Finally, we explain the microscopic origin of these phenomena by constructing and analyzing effective low-energy models.

This paper is organized as follows:
In Sec.~\ref{Section::Model}, we introduce the model Hamiltonian whose ground state is found by DMRG simulations.
In Sec.~\ref{Sction::Results}, we calculate observables such as the magnetization, the winding number, the entanglement distribution and the spin-spin correlation functions in order to differentiate between SkX and SkL phases.
Additionally, we show that a purely classical treatment of the spin chain does not host a SkL phase.
In Sec.~\ref{Section::MicroscopicDescription}, we use a microscopic model to argue that the SkL phase is a result of a magnetic BEC.
Moreover, we demonstrate that superpositions of translated classical skyrmions can approximate the entanglement distribution of quantum skyrmions in the SkL phase.
Finally, we conclude and summarize our results in Sec.~\ref{Section::Summary}.

\section{\label{Section::Model}Model and Method for numerics}
We use the following spin Hamiltonian on a triangular lattice,
\begin{equation}
    \pazocal{H}=\sum_{\left\langle\boldsymbol{r}, \boldsymbol{r}'\right\rangle} 
    \left[ J \hat{\boldsymbol{S}}_{\boldsymbol{r}} 
    \cdot 
    \hat{\boldsymbol{S}}_{\boldsymbol{r}'} + \boldsymbol{D}_{\boldsymbol{r}-\boldsymbol{r}'} 
    \cdot \left(\hat{\boldsymbol{S}}_{\boldsymbol{r}}\times \hat{\boldsymbol{S}}_{\boldsymbol{r}'}\right) \right]
    -
    \sum_{\boldsymbol{r}} B_z \hat{S}_{\boldsymbol{r}}^z,
    \label{eq::System_Hamiltonian}
\end{equation}
where $\hat{\boldsymbol{S}}_{\boldsymbol{r}}$ are $S=1/2$ operators at the lattice site at position $\boldsymbol{r}$.
Throughout this article, we consider a ferromagnetic Heisenberg exchange interactions with strength $J<0$.
$\boldsymbol{D}_{\boldsymbol{r}-\boldsymbol{r}'}$ is the amplitude of the in-plane Dzyaloshinskii-Moriya interaction (DMI) between sites $\boldsymbol{r}$ and $\boldsymbol{r}'$, and is given by $\boldsymbol{D}_{\boldsymbol{r}-\boldsymbol{r}'}=D\hat{e}_z\times(\boldsymbol{r}-\boldsymbol{r}')$ with DMI strength $D$. 
Moreover, $B_z$ represents an external magnetic field perpendicular to the two-dimensional lattice plane.
Additionally, the summation over $\left\langle\boldsymbol{r}, \boldsymbol{r}'\right\rangle$ is over all unique nearest-neighbor bonds.
For convenience we fix the parameters $J=-0.5$ and $D=1$ throughout this paper while $B_z$ is variable.
Moreover, without loss of generality, we choose to work with a triangular lattice for the numerical simulations.
The system size is fixed to $L_x=251$, $L_y=9$, where $L_x$ and $L_y$ are the number of sites along the primitive directions $\hat{e}_x$ and $(1/2)\hat{e}_x+(\sqrt{3}/2)\hat{e}_y$, respectively.

\begin{figure*}[ht]
\centering
\includegraphics[width=1.0\textwidth]{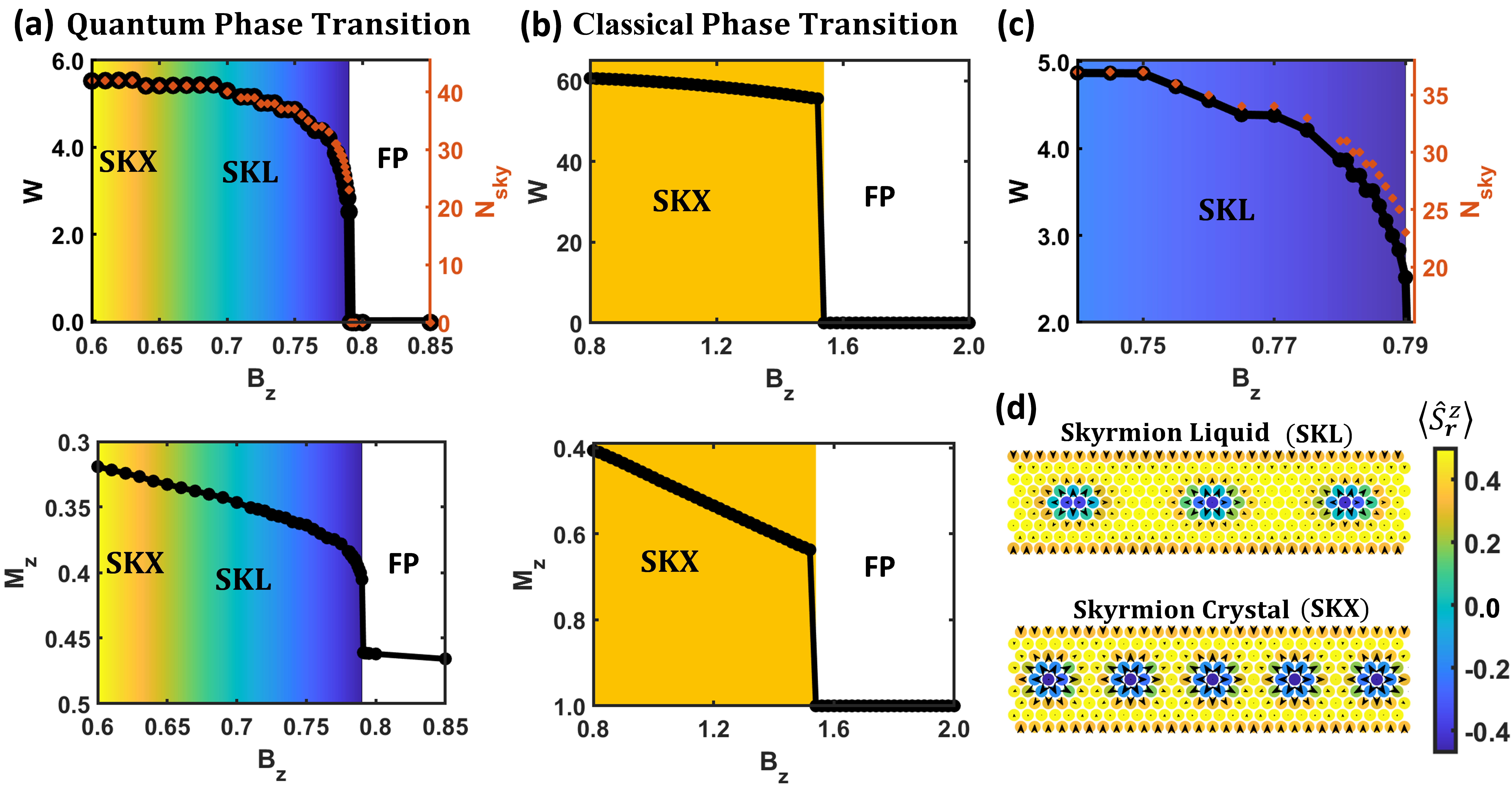} 
\caption{(a) The winding number (top) and magnetization (bottom) as a function of the applied magnetic field for a spin-half quantum system. The similarity between the number of skyrmions (see the pink dots) and the winding number as a function of the magnetic field confirms that the winding number is proportional to the number of skyrmions in the system. The gradual change in color from yellow to blue denotes the crossover from SkX to SkL phase. (b) The winding number (top) and magnetization (bottom) as a function of magnetic field for a classical spin system with spin of unit magnitude at temperature $T=0.001D$. (c) The deviation between the winding number $W$ and the number of quantum skyrmions $N_{sky}$ is dominant near the phase transition (SkL regime) indicating that the skyrmion nature is more quantum near the transition point. (d) Real-space spin configuration of the system in the skyrmion liquid phase (top) and skyrmion crystal phase (bottom).
}
\label{fig::PhaseTransition}
\end{figure*}

This system has been recently investigated successfully using DMRG~\cite{AndreasHaller}, which is an established numerical technique for exploring exotic phases in quasi-one-dimensional systems~\cite{DMRG_Example1, DMRG_Example2, DMRG_Example3, DMRG_Example4, DMRG_Example5, DMRG_Example6, DMRG_Example7, DMRG_Example8, DMRG_Example9, DMRG_Example10}.
However, since DMRG is a variational technique, the algorithm tends to get stuck in local minima for large system sizes, potentially leading to spurious results.
This outcome can be avoided by running DMRG with multiple initial states with different numbers of classical skyrmions (which can be represented by a matrix product state with bond dimension $\chi=1$) and collecting the results only for the lowest energy state.
Unless otherwise specified, the bond dimension of the matrix product states we use in our simulations is fixed at $\chi=150$.
This bond dimension is large enough to ensure well-converged expectation values for energies and local order parameters.
The numerical results are further presented in the next section.

\section{\label{Sction::Results} Numerical Results}

In this section, we characterize three phases (SkX, SkL, FP) of the system by studying different order parameters and the entanglement distribution throughout the system. The winding number for a triangular lattice spin system is,
\begin{align}
    W&=\frac{1}{2\pi}\sum\limits_{i,j,k\in\Delta/\nabla} \arctan\left(x_{ijk} \right) \notag \\
    x_{ijk} &=
    \frac
    {8\left\langle \boldsymbol{S}_i\cdot (\boldsymbol{S}_j\times\boldsymbol{S}_k)\right\rangle}
    {1+4\left(
    \left\langle \boldsymbol{S}_i\cdot\boldsymbol{S}_j\right\rangle
    +\left\langle \boldsymbol{S}_i\cdot\boldsymbol{S}_k\right\rangle
    +\left\langle \boldsymbol{S}_j\cdot\boldsymbol{S}_k\right\rangle\right)},
\end{align}
where $i,j,k$ denote the sites, in counterclockwise order, of triangles pointing up ($\Delta$) or down ($\nabla$).
This winding number evaluates the solid angle spanned by all the skyrmion spin configurations divided by $4\pi$.
In the classical continuum limit, this is directly related to the total number of skyrmions in the system.
The winding number as a function of the magnetic field is shown in Fig.~\ref{fig::PhaseTransition}(a).
As the magnetic field increases the winding number decreases continuously, which denotes a smaller number of skyrmions in the system at a higher magnetic field.
The orange symbols in Fig.~\ref{fig::PhaseTransition}(a) demonstrate that the winding number $W$ is, to a large extent, proportional to the number of skyrmions $N_{\rm sky}$ in the system, which are obtained by counting the number of negative $S^z_{\boldsymbol{\bm r}}$ domains (blue colored domains in Fig.\ref{fig::PhaseTransition}(d)) along the ribbon center.
The gradual change in color in Fig.\,\ref{fig::PhaseTransition}(a) indicates a crossover from the skyrmion crystal (SkX) to the skyrmion liquid (SkL) phase.
The distinction between the SkX and SkL phases lies in the density of the skyrmions (Fig.~\ref{fig::PhaseTransition}(d)). 
In the SkX phase, the skyrmions are densely packed, whereas in the SkL phase, they are more loosely packed.
At magnetic field $B>0.79D$, the system becomes fully polarized.
Fig.\,\ref{fig::PhaseTransition}(c) illustrates the discrepancy in the number of skyrmions and the winding number near the phase transition point, which arises due to the quantum nature of skyrmions.

Furthermore, a plot of the magnetization $M_z$ as a function of the magnetic field in Fig.~\ref{fig::PhaseTransition}(a) confirms that the quantum phase transition is second order in nature.
We simulated the same system using classical Monte Carlo and show both the winding number as well as magnetization as a function of the magnetic field in Fig.~\ref{fig::PhaseTransition}(b).
In stark contrast to the quantum regime, the phase transition of the classical spin system at zero temperature turns out to be first-order in nature and contains no SkL phase.
This result is compatible with the reports of Yoshihiko \textit{et al.}~in Ref.~\cite{ClassicalSupportingResult}.
The second-order phase transition and the presence of the SkL must therefore be caused by strong quantum spin fluctuations.

As we will elaborate in more detail in Sec.~\ref{Section::MicroscopicDescription}, these phenomena can be explained as follows by quantum skyrmion band structures.
In the FP regime, a single skyrmion is an excitation above the polarized ground state and forms a skyrmion band structure due to quantum skyrmion-lattice interaction. 
The bandwidth of the band structure is a measure of the kinetic energy of the skyrmion.
As the magnetic field decreases past the critical point, the energy of the skyrmionic band minima crosses the energy of FP phase, favoring the formation of skyrmions in the system.
However, to compensate for the kinetic energy of skyrmions, we have fewer of them instead of a saturated number.
In stark contrast, in the low-temperature classical spin limit, the flat skyrmion bands lead to zero skyrmionic kinetic energy. Thereby favoring the presence of the maximum number of skyrmions in the system due to a lack of kinetic energy.
Thus reduces the order of the phase transition by one.

It is important to note that the two different results in Fig.~\ref{fig::PhaseTransition} are obtained from quantum simulations of $S=1/2$ systems and a fully classical treatment of spins with unit magnitude.
Quantum simulations of the same Hamiltonian \eqref{eq::System_Hamiltonian} for higher spin values $S$ might not lead to a tendency towards a first-order phase transition because, as we discuss in App.~\ref{Appendix::A}), increasing the spin value does not necessarily increase the ``classicality'' of a skyrmion:

The stabilization of a skyrmion phase requires a finite magnetic field, and the critical value of the magnetic field increases with the spin value.
Since a larger magnetic field also reduces the radius of the skyrmion, higher spin values lead to a reduction in the skyrmion radius, which in turn increases the relevance of the background lattice discretization and thus the associated quantum fluctuations.

For example, in the limiting case of a classical spin system ($S\rightarrow\infty$) an infinite magnetic field ($B_z\rightarrow\infty$) is required to stabilize a skyrmion phase.
This would result in a zero skyrmion radius ($R\rightarrow 0$) and would increase quantum fluctuations.

Due to this intricate interplay between stabilizing magnetic field, the background lattice and the spin, we see qualitatively the same change of the number of skyrmions versus renormalized Zeeman field close to the phase transition between skyrmion liquid and polarized phase (see Fig.\,\ref{fig::QuantumPhaseTransition_DifferentSpins}), which indicates that the skyrmion liquid phase is not limited to the special case $S=1/2$ and is likely present in systems with larger spin values. In Fig.\,\ref{fig::PhaseTransition}, we therefore present the two extreme cases of $S=1/2$ and classical spins.

\begin{figure*}[ht]
\centering
\includegraphics[width=1.0\textwidth]{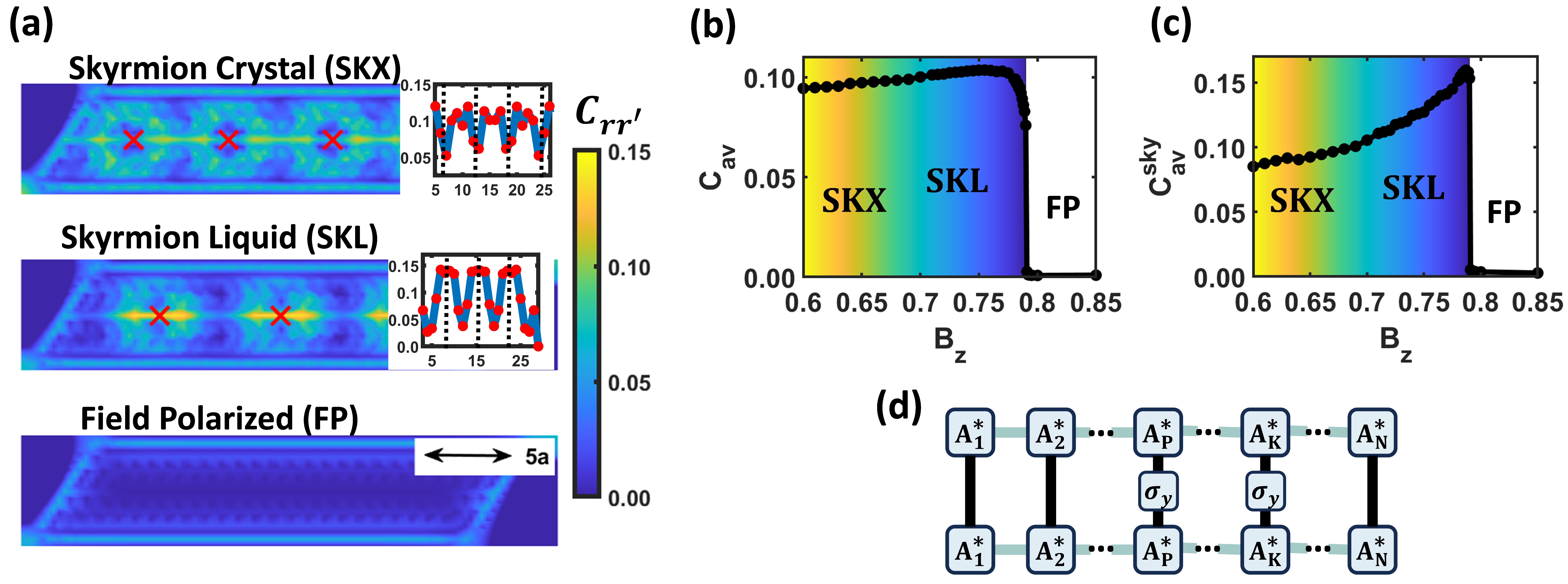} 
\caption{Entanglement distribution (concurrence) in the three different phases. The skyrmion centers are denoted by red crosses. The insets show the concurrence plotted along a line connecting the skyrmion centers and the skyrmion centers are denoted by vertical dotted lines. The plot shows a significant difference in the entanglement distribution between the SkX and SkL phases. 
(b) Average concurrence per nearest-neighbor bond as a function of magnetic field.
(c) Concurrence averaged over the nearest-neighbor bonds near the skyrmion center as a function of the magnetic field. It shows that the concurrence is localized near the skyrmion center in the SkL phase, whereas it decays and moves outwards in the SkX phase.
(d) A schematic representation of concurrence calculation using the matrix product state representation. Here, the concurrence is calculated on a bond between $P$th and $K$th sites. The tensors $(A_1, A_2, \cdots, A_N)$ form the matrix product representation of the wavefunction.
}
\label{fig::EntanglementDistribution}
\end{figure*}

Next, we study the entanglement distribution throughout the system, for which we calculated the concurrence of nearest-neighbor spin pairs.
The concurrence is defined as\,\cite{Wootters1998, AndreasHaller},
\begin{align}
    C_{\boldsymbol{r}\boldsymbol{r}'} = \big|\bra\psi \hat F_{\boldsymbol{r}\boldsymbol{r}'}\left(\ket\psi^*\right)\big|
    ,\quad
    \hat F_{\boldsymbol{r}\boldsymbol{r}'} = \sigma^{y}_{\boldsymbol{r}}\sigma^y_{\boldsymbol{r}'}
    \label{eq:concurrence_definition}
\end{align}
where $\hat F_{\boldsymbol{r}\boldsymbol{r}'}$ is the spin flip operator associated with the spins at positions $\boldsymbol{r}$ and $\boldsymbol{r}'$. A schematic for the concurrence calculation based on matrix product states is shown in Fig.~\ref{fig::EntanglementDistribution}(d).
We would like to emphasize the conjugate ket state $\ket{\psi}^*$ in the definition of $C_{\boldsymbol{r}\boldsymbol{r}'}$, which drastically changes the properties of the concurrence from those of an ordinary two-point correlation function.
$C_{\boldsymbol{r}\boldsymbol{r}'}\in[0,1]$ is a bounded function that can measure the entanglement between two spins at sites $\boldsymbol{r}$ and $\boldsymbol{r}'$.
The concurrence is related to the entanglement of formation~\cite{Wootters1998} which quantifies the quantum correlations also for spin pairs that cannot be represented by pure states.
In our system, we use the concurrence to measure the local entanglement between two nearest-neighbor spins in the lattice.
In Fig.~\ref{fig::EntanglementDistribution}(a), it is observed that the presence of skyrmions creates a nontrivial entanglement distribution throughout the system whereas the entanglement remains small and uniform in the FP phase.
However, the entanglement distribution of the states in the SkL and SkX phases is markedly different.
The entanglement in the SkX states is mainly distributed around the skyrmion boundary.
In the SkL states, in contrast, the entanglement is more concentrated near the skyrmion center.
We argue below that this behavior is caused by the uncertainty of the skyrmion position and its concomitant quantum fluctuations, which in turn can be noticed in the real-space spin configuration of skyrmions in the SkL states, see Fig.~\ref{fig::PhaseTransition}(c).
Moreover, the uncertainty of the skyrmion position can be interpreted as resulting from a superposition of many quantum skyrmions located around the classical skyrmion center, which is why the entanglement is highly concentrated near the skyrmion center (see Sec.~\ref{Section::MicroscopicDescription} and Fig.~\ref{fig::EntanglementDistribution_MicroscopicModel} for more details).

Furthermore, in Figs.~\ref{fig::EntanglementDistribution}(b) and (c) we have plotted, respectively, the concurrence $C_{\rm av}$ averaged over the whole system and the concurrence $C_{\rm av}^{\rm sky}$ averaged over the bonds near the skyrmion center as a function of the magnetic field.
This shows that the average entanglement per bond of the system does not vary much in the SkX and SkL phases, but the entanglement distribution gradually moves towards the skyrmion center in the SkL phase.
This continuous change in entanglement distribution is also associated with the continuous change in the number of skyrmions from the SkX to SkL phases.
As the skyrmions are packed more closely in the SkX phase, the uncertainty or quantum fluctuation of the skyrmion position is lost due to skyrmion-skymion interaction.
As a result, the skyrmions in the SkX phase are not composed of a superposition of many skyrmionic states centered around the classical skyrmion position, and thus the entanglement at the skyrmion center disappears.
Shifting entanglement towards the boundary is a clear indication of the strong correlation of skyrmions in the SkL phase.

\begin{figure}[tb]
\centering
\includegraphics[width=0.5\textwidth]{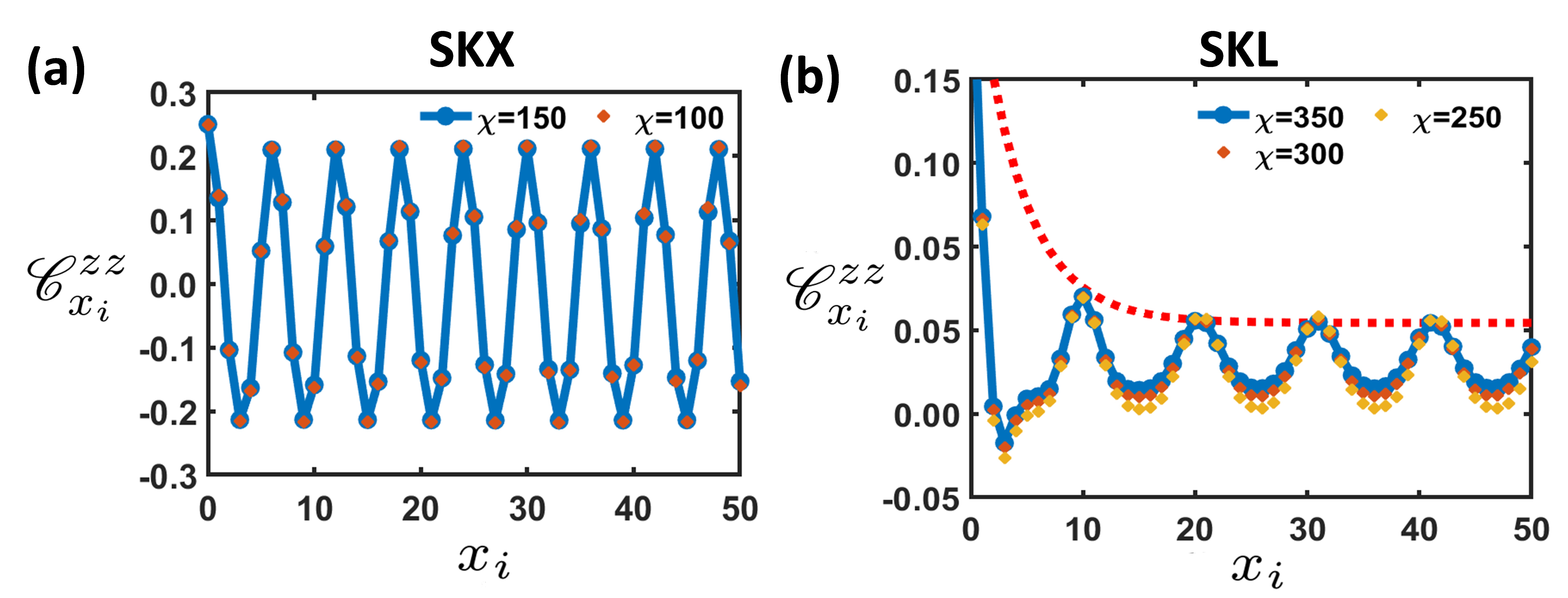} \caption{(a) Spin-spin correlation $\mathcal{C}^{zz}_{x_i}$ as a function of distance $x_i$ from one of the skyrmion centers is plotted for (a) skyrmion crystal phase at $B_z=0.62$ and (b) skyrmion liquid phase at $B_z=0.789$. The inset in (b) shows the fit of the correlation peaks with an exponential function $A\ \text{exp}(-Bx_i^{\text{peak}})+C$ with parameters $A=0.2454$, $B=0.2499$, $C=0.004575$. This shows that the skyrmion-skyrmion correlation is long-ranged (short-ranged) in the SkX (SkL) phase.
}
\label{fig::SpinSpinCorrelation}
\end{figure}

We have calculated the following correlation function, 
\begin{equation}
\mathcal{C}^{zz}_{x_i}=\left\langle\hat{S}^z_{X_0+x_i,Y_0} \hat{S}^z_{X_0,Y_0}\right\rangle,
\end{equation}
where the coordinate $(X_0, Y_0)$ is at the center of one of the skyrmions and $x_i=ia$ ($i \in \mathbb{Z}$) is the distance along the horizontal axis.
In Fig.~\ref{fig::SpinSpinCorrelation}(a) and (b) we plot, respectively, the correlation function $\mathcal{C}^{zz}_{x_i}$ as a function of distance from the skyrmion peak for the SkX and SkL ground states.
The numerical calculation of $\mathcal{C}^{zz}_{x_i}$ converges fast in the SkX phase already for smaller bond dimensions, whereas the ground states in the SkL phase require a higher bond dimensions for better convergence.
For $x_i \to 0$, the correlations converge to $\mathcal{C}^{zz}_{x_i=0}=1/4$ because $(S^z_i)^2=1/4$ for spin-$1/2$ sites.
The correlations between nearest-neighbor skyrmion centers are lower than the quantized on-site value and the difference between SkL and SkX states is substantial.
In addition, at larger distances the correlation function remains periodic in the SkX phase but decays in the SkL phase.
This indicates that the quantum skyrmions in the SkX phase are spatially strongly correlated.
In contrast, we find that the spin-spin correlation function in the SkL state decays with distance, which we attribute to a fluid-like behavior of the uncorrelated skyrmions.

In conclusion, we claim that the SkL phase which has a very distinct characteristic compared with the conventional SkX phase is a consequence of the kinetic energy of skyrmions originating due to skyrmion-lattice interaction.
Although this kinetic energy is the key component for the realization Bose-Einstein condensate of skyrmions, in our study of skyrmions in a quasi-one-dimensional system, it does not lead to a BEC.
This absence of BEC is essentially due to the hardcore nature of skyrmions.
A BEC for hardcore bosons can only be realized if next-nearest hopping is allowed which eventually allows the bosonic wavefunction to be evenly distributed over the system\,\cite{HardCoreBoson}.
Thus, we assert that even though we have not achieved BEC in this simple system, our study encourages further exploration in this area.
Particularly, we propose that a system with a similar Hamiltonian as ours, but with higher dimension or with artificially shifted boundary conditions, has a possibility of having BEC of skyrmions.

\section{\label{Section::MicroscopicDescription} Microscopic Description}

We use the collective coordinate formalism\,\cite{CollectiveCoordinate1, CollectiveCoordinate2, Tserkovnyak, LeonBalents} and trial wave functions to explain the numerical results obtained in the previous section.
The excited state of the FP phase hosts delocalized skyrmions, which are formed by the modes of a dispersive band structure of single skyrmions due to their interaction with the lattice (see Fig.~\ref{fig::Schematic}).
As the magnetic field is lowered, the skyrmionic excited state becomes the ground state.
However, the system fails to achieve saturated numbers of skyrmions due to the kinetic energy of skyrmions, resulting in the formation of the SkL phase.
The SkX phase is formed as the magnetic field, which acts as a chemical potential for skyrmions, is lowered.
In this line of reasoning, a dispersive single-skyrmion band is essential to form a SkL phase.
In this section, we demonstrate how the skyrmion size and the value of the microscopic spin affect the band dispersion.

\begin{figure*}[ht]
\includegraphics[width=0.9\textwidth]{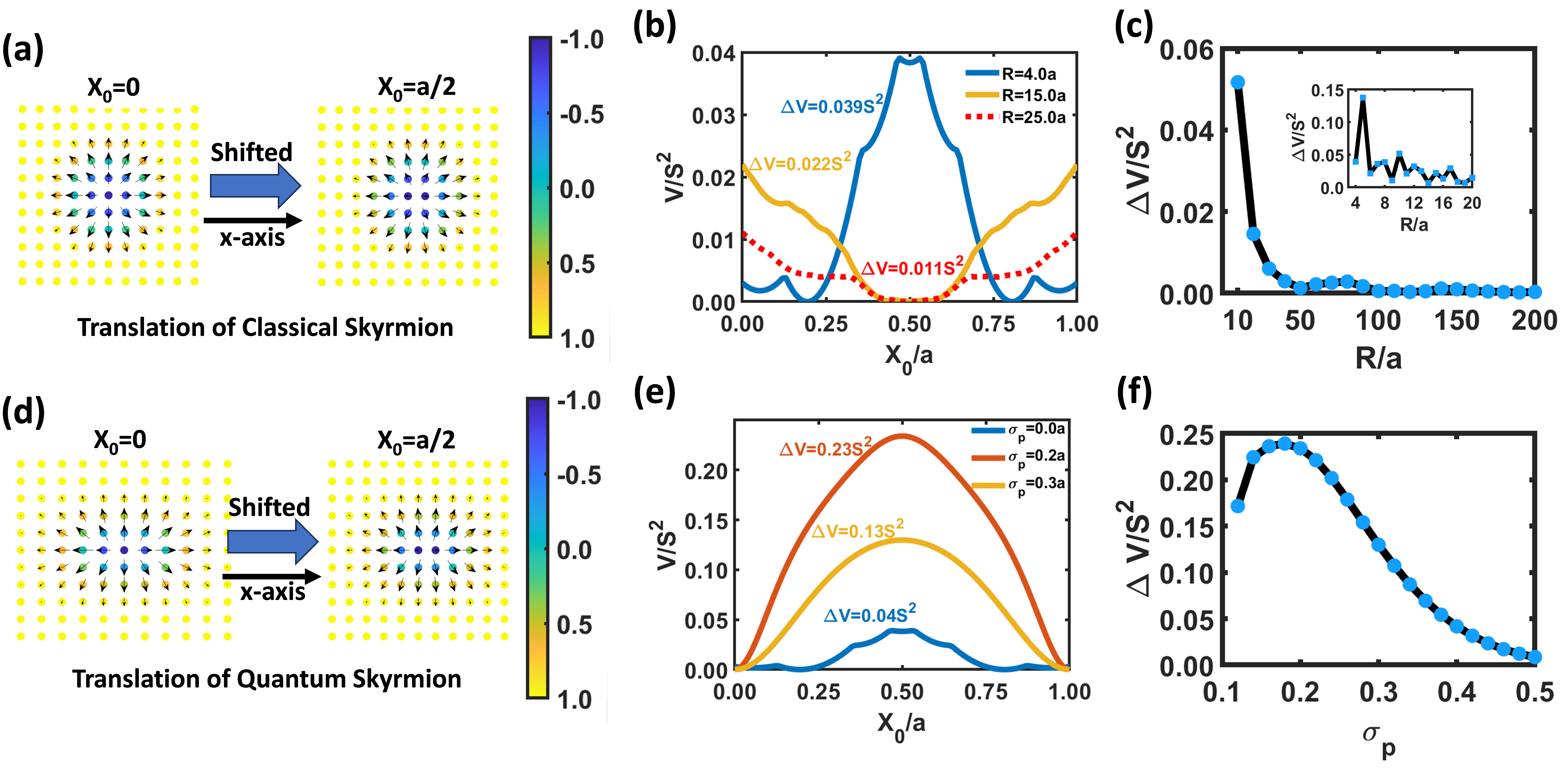} 
\caption{(a) Schematic for the translation of a classical skyrmion over a lattice. 
(b) Potential profile for classical skyrmions of different radii, showing that the width of the potential decreases as the radius increases.
(c) Potential width as a function of the radius of a classical skyrmion. The inset shows a magnified plot to show the roughness of the potential width as a function of radius.
(d) Schematic for the translation of a quantum skyrmion over a lattice.
(e) Potential profile for a quantum skyrmion for different $\sigma_p$ (quantum fluctuation of skyrmion position) with radius $R=4a$.
(f) Potential width as a function of $\sigma_p$ with radius $R=4a$. The potential width does not vary monotonically with the uncertainty of the skyrmion position.
}
\label{fig::Skyrmion_Potential_Profile}
\end{figure*}

We now explain how dispersive bands result from the skyrmion-lattice interaction.
Instead of behaving like a term with a tendency to confine a skyrmion with wavefunction $\Psi_{\pazocal{S}}(\boldsymbol{X})$ at the potential minima $V(\boldsymbol{\hat{X}}) \Psi_{\pazocal{S}}(\boldsymbol{X})=V(\boldsymbol{{X}}) \Psi_{\pazocal{S}}(\boldsymbol{X})$, the potential operator $V(\boldsymbol{\hat{X}})$ acts as a translational operator on a skyrmionic wavefunction, $V(\boldsymbol{\hat{X}}) \Psi_{\pazocal{S}}(\boldsymbol{X})=\sum_{\boldsymbol{T}} C_{\boldsymbol{T}} \Psi_{\pazocal{S}}(\boldsymbol{X}+\boldsymbol{T})$, due to non-trivial commutation relation between the components of skyrmion center $\hat{X}_0$ and $\hat{Y}_0$ as discussed later in this section.
Thus, skyrmion band structures are more dispersed when the potential profile is more dispersed due to stronger skyrmion-lattice interaction.
As skyrmions grow larger, the lattice appears to be more continuous, and as a consequence, the skyrmion-lattice interaction should become negligible.
To demonstrate the presence of such a potential, we use a classical skyrmion configuration defined by the spin configuration $(\sin(\theta_s) \cos(\phi_s), \sin(\theta_s)\sin(\phi_s), \cos(\theta_s) )$, where the spherical-polar angles are functions of position given by
\begin{align}
    \theta_s&=\frac{\pi r}{R}, & \phi_s &= \tan^{-1}\left( \frac{y-Y_0}{x-X_0}\right) & \text{for } r<R,\nonumber\\
    \theta_s&=\pi,& \phi_s& =0, &\text{for } r\geq R,
    \label{eq::Classical_Skyrmion_Config}
\end{align}
where $r=\sqrt{(x-X_0)^2+(y-Y_0)^2}$ and $R$ is the skyrmion radius.
Here, $(x,y)$ and $(X_0,Y_0)$ denote the coordinate of a lattice site and the skyrmion center, respectively.
The $y$ component of the skyrmion center $Y_0$ is fixed to $(L^y+1)/2$, where $L^y$ is an odd number denoting the length of the system in the $y$-direction.
The potential as a function of skyrmion position is given by
\begin{equation}
V(X_0)=E(X_0)-E_{\text{fp}},
\label{eq::Skyrmion_Potential}
\end{equation}
where $E(X_0)$ is the energy of the whole system in the presence of a skyrmion, whereas $E_{\text{fp}}$ is the energy of the system in the absence of skyrmions. 
Through the translation of the skyrmion configuration as in Fig.\,\ref{fig::Skyrmion_Potential_Profile}(a), we have assessed the potential profile for different skyrmion radii, see Fig.\,\ref{fig::Skyrmion_Potential_Profile}(b). 
The potential width decreases with increasing skyrmion radius, validating our claim that the skyrmion-lattice interaction diminishes for large radii as in Fig.~\ref{fig::Skyrmion_Potential_Profile}(c).
Until now, we have considered the potential profiles of classical skyrmions.

A classical skyrmion is equivalent to a product state, because it can be written in terms of local rotations of a fully-polarized state $\left|\Uparrow\right\rangle$ as,
\begin{equation}
    \left|\Psi_{\pazocal{S}}^R(X_0)\right>=\prod_{\boldsymbol{r}} \exp(-i\phi_s(\boldsymbol{r})\hat{S}^z_{\boldsymbol{r}}) \exp(-i\theta_s(\boldsymbol{r})\hat{S}^y_{\boldsymbol{r}}) \left|\Uparrow\right\rangle,
\end{equation}
where $\theta_s(\boldsymbol{r})$ and $\phi_s(\boldsymbol{r})$ follow the expressions in Eq.~\eqref{eq::Classical_Skyrmion_Config}.
Although the skyrmion profile is now formally represented by a quantum state, the corresponding potential profile is the same as the one for a classical skyrmion because the state is devoid of entanglement.

To simulate a quantum skyrmion, we need to consider quantum spin fluctuations of the position of the skyrmion center, which is characterized by the non-commutativity,
\begin{equation}
    \left[\hat{X}_0, \hat{Y}_0\right]=il_N^2,
    \label{eq::Skyrmion_Commutation_Relation}
\end{equation}
where the magnetic length is $l_N=\sqrt{{A_c}/{4\pi SQ}}$.

In order to account for such quantum fluctuations, we have constructed an MPS representation of a quantum skyrmion by superposing many skyrmions with different centers with a Gaussian probability distribution,
\begin{align}
&
    \ket{\Psi_{\pazocal{S}}^{R,\sigma_p}(X_0)}  \notag \\
&=\frac{1}{\sqrt{N}}
    \sum_{n\in \mathbb{Z}} 
    \exp[-\frac{1}{2}\left(\frac{n\Delta X_0}{\sigma_p}\right)^2]
    \ket{\Psi_{\pazocal{S}}^R(X_0+n\Delta X_0)},
    \label{eq::SkyrmionicWavefunction}
\end{align}
where numerically we fixed $\Delta X_0=0.0005 a$ and chose the maximum of $|n|$ such that $|n|\Delta X_0<7\sigma_{p}$.
The resulting wavefunction converges to a continuum limit wavefunction ($\Delta X_0 \rightarrow 0$) as we observed that the potential profile saturates by tuning $\Delta X_0$ more finely.
We construct the wavefunction $\Psi_{\pazocal{S}}^{R,\sigma_P}$ at various positions $X_0$ on the lattice (see Fig.~\ref{fig::Skyrmion_Potential_Profile}(d)) and then evaluate the skyrmion potential profile using Eq.~\eqref{eq::Skyrmion_Potential}.
In Fig.~\ref{fig::Skyrmion_Potential_Profile}(e) we see that although the potential width in the classical limit $\sigma_p \to 0$ is lower than for the quantum skyrmion at $\sigma_p=0.2a$, a further increase in $\sigma_p$ to $0.3a$ leads again to a decrease in potential width.
Thus the potential width does not vary monotonically with $\sigma_p$ as is shown in Fig.~\ref{fig::Skyrmion_Potential_Profile}(f).

\begin{figure*}[ht]
\centering
\includegraphics[width=0.85\textwidth]{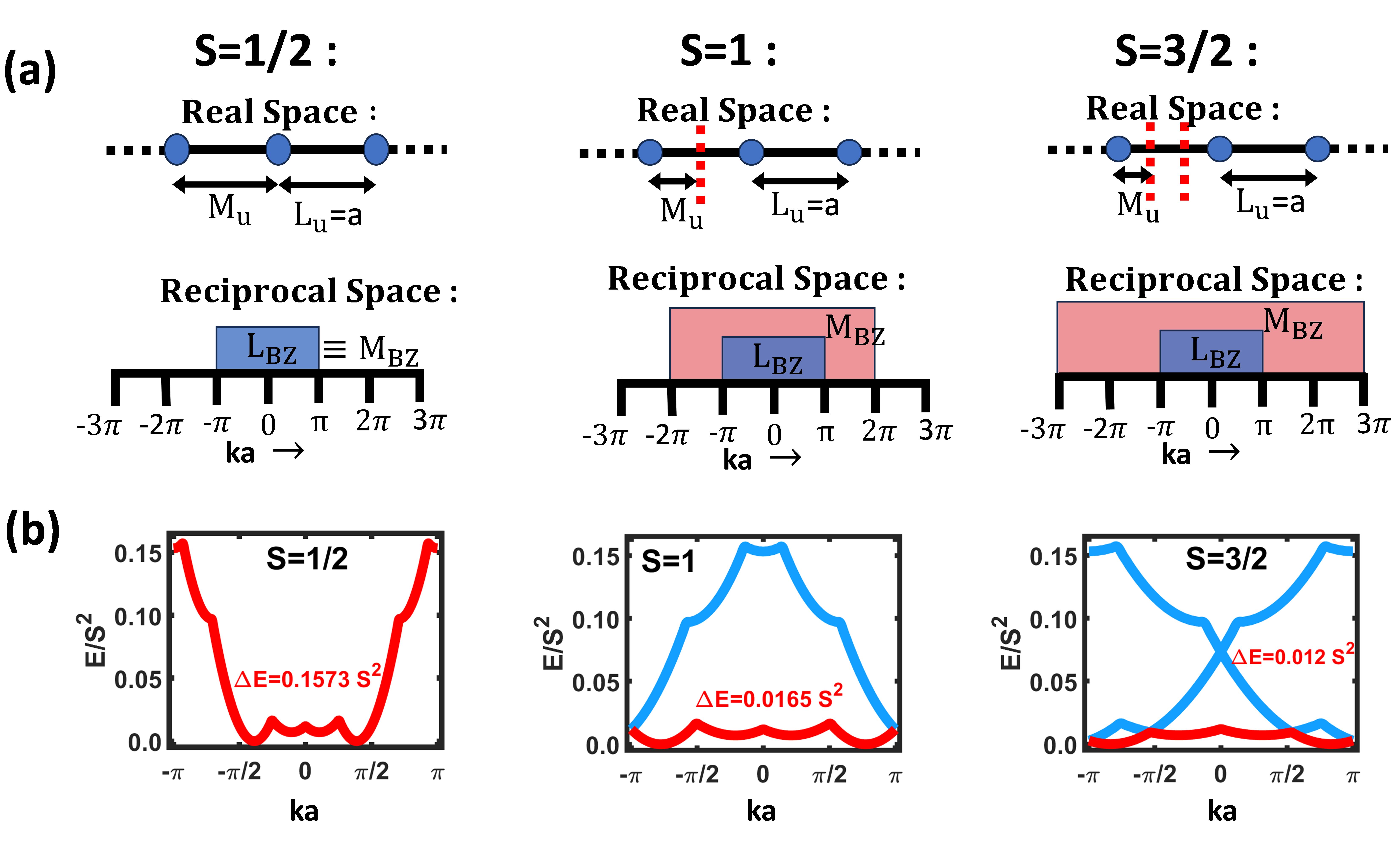} 
\caption{(a) (Top) The magnetic ($\text{M}_\text{u}$) and lattice ($\text{L}_\text{u}$) unit cells are depicted for spins $S=1/2$, $1$, $3/2$. (Bottom) The corresponding lattice ($\text{L}_{\text{BZ}}$) and magnetic ($\text{M}_{\text{BZ}}$) Brillouin zones are shown.  (b) The skyrmion band structures are plotted within the lattice Brillouin zone ($\text{L}_{\text{BZ}}$) for different spins $S=1/2$, $1$, $3/2$, and the lowest band is shown in red. For higher spin, the magnetic Brillouin zone is larger and thus multiple band folding occurs at the lattice Brillouin zone boundary. The resulting individual bands have lower bandwidth and only the bandwidth of the lowest band is shown in red text.}
\label{fig::Skyrmion_Band_Structure}
\end{figure*}

Although the potential depth decreases for larger skyrmions, it does not vanish completely in the classical limit. Thus, the absence of a classical SkL phase cannot be ruled out only based on the potential depth.
To explain the phenomenon we need to take into account the quantum magnetic flux experienced by a skyrmion due to its non-trivial spin texture.
The information on the quantum flux is embedded in the commutation relation Eq.~\eqref{eq::Skyrmion_Commutation_Relation}, which is equivalent to the commutation relation for the guiding center of an electron in a magnetic field,
\begin{equation}
    \left[\hat{X}_e, \hat{Y}_e\right]=il_B^2,
\end{equation}
where $l_B=\sqrt{\hbar c/eB}$ is the magnetic length for a magnetic field $B$.
Thus, in this collective coordinate formalism, the only difference between the two systems is in the scaling of the magnetic length.
For the electronic system $l_B\gg a$, whereas for the skyrmionic system $l_N \leq a$, which inverts the scaling between magnetic and lattice unit cell.
Due to the non-trivial commutation relations in Eq.~\eqref{eq::Skyrmion_Commutation_Relation}, the translation operators of the system have the properties,
\begin{align}
    \hat{T}_{x}=\text{e}^{-i\frac{2\pi}{a}\hat{Y}_0}&,\,\,\,
    \hat{T}_{y}=\text{e}^{i\frac{2\pi}{a}\hat{X}_0},
    \nonumber\\
    \hat{T}_{x}^{\dagger} \hat{X}_0 \hat{T}_{x} = \hat{X}_0+\frac{a}{N}&,\,\,\,
    \hat{T}_{y}^{\dagger} \hat{Y}_0 \hat{T}_{y} = \hat{Y}_0+\frac{a}{N}.
\end{align}
The operators $\hat{T}_x$ and $(\hat{T}_x)^N$ correspond to translations by one magnetic and lattice unit cell, respectively.
As a consequence, the relationship between the sizes of the magnetic ($M_u$) and lattice unit cells ($L_u$) becomes $M_u=L_u/N$ which is shown in Fig.~\ref{fig::Skyrmion_Band_Structure}(a).
Moreover, the Bloch states $\ket{k}$, which are the eigenstates of the lattice translation operator,
\begin{equation}
(\hat{T}_x)^N\ket{k}=\text{e}^{ika}\ket{k},    
\label{eq::Bloch_State1}
\end{equation}
are no longer eigenstates of $\hat{T}_x$, 
\begin{equation}
    \hat{T}_x\ket{k}=\text{e}^{\frac{ika}{N}} \ket{k+\frac{2\pi}{a}}
    \label{eq::Bloch_State2}
\end{equation}
and thus $\ket{k+n2\pi/a}$ is no longer equivalent to $\ket{k}$ (where $n\in\mathbb{Z}_N$).
However, the states $\ket{k}$ and $\ket{k+2\pi N/a}$ are equivalent to each other (see Eqs.~\eqref{eq::Bloch_State1} and \eqref{eq::Bloch_State2}), increasing the size of the Brillouin zone to $2\pi N/a$, which defines the magnetic Brillouin zone.
As a consequence, the magnetic Brillouin zone is larger than the lattice Brillouin zone, and sizes are related as $M_{\text{BZ}}=NL_{\text{BZ}}$ (see Fig.\,\ref{fig::Skyrmion_Band_Structure}(a)).

However, band folding is necessary to bring back the bands into the lattice Brillouin zone.
This is done by defining a new quantum number $q\in\mathbb{Z}_N$ such that $\ket{k,q}=\ket{k+(q-1)2\pi/a}$.
The Hamiltonian for the skyrmion in the collective coordinate formalism is $\hat{H}=\sum_n V_n (\hat{T}_x)^n$, where $V_n$ is the Fourier transformed periodic lattice potential experienced by the skyrmion.
By expanding the Hamiltonian in the basis $\ket{k,q}$ we obtain the components of the Hamiltonian matrix as,
\begin{equation}
    \hat{H}_{\alpha\beta}(k)=
    \left(\sum_{p\in\mathbb{Z}} V_{\alpha-\beta+pN} \text{e}^{ipka}\right)
    \text{e}^{\frac{i(\alpha-\beta)ka}{N}}.
\end{equation}
By diagonalizing this matrix we obtain the skyrmion band structure.
In Fig.\,\ref{fig::Skyrmion_Band_Structure}(b), we show the skyrmion band structures for different spins $S \in \{ 1/2, 1, 3/2 \}$ using the potential in Fig.~\eqref{fig::Skyrmion_Potential_Profile}(b) for $\sigma_p=0$.
We use the bandwidth of the lowest band (denoted by red color) to estimate the average kinetic energy of skyrmions.
We observe that the bandwidth of the lowest skyrmion band decreases with spin due to the band folding which originated from the mismatch between lattice and magnetic unit cell (or Brillouin zone) in the artificial gauge field experienced by the skyrmion due to its topological spin-texture.
Thus in the classical limit $S\rightarrow \infty$, we expect the bandwidth of the bands $\Delta E\rightarrow 0$ due to an infinite numbers of band foldings, denoting the absence of skyrmion kinetic energy.
As a consequence, an infinitesimal skyrmion-skyrmion interaction can destroy the SkL phase in the classical limit, which explains the absence of the SkL phase between SkX and FP in Fig.~\ref{fig::PhaseTransition}. 
A short discussing of skyrmion-skyrmion interaction is provided in App.~\ref{Appendix::B}.

\begin{figure}[tb]
\centering
\includegraphics[width=0.45\textwidth]{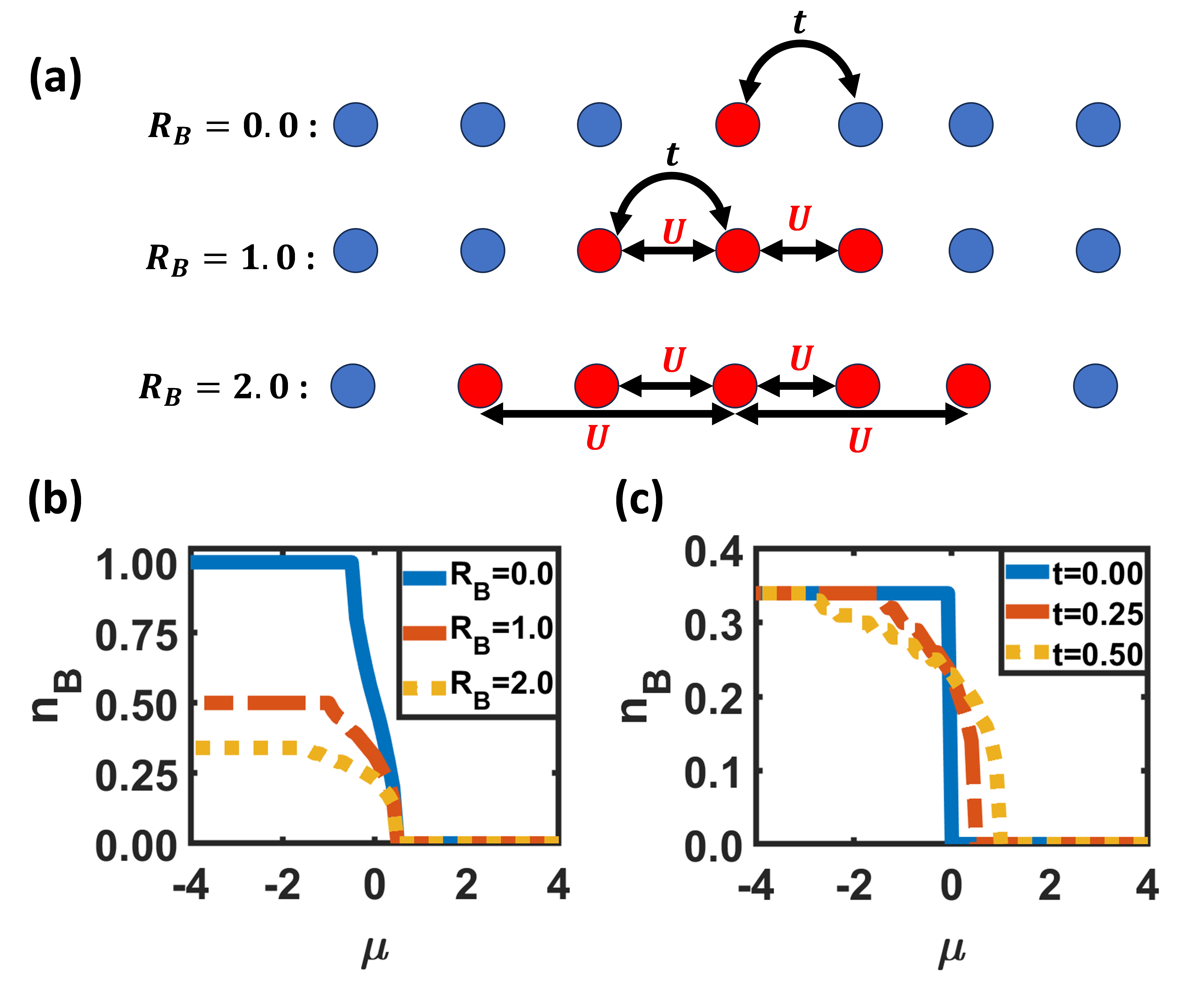} 
\caption{ (a) Schematic picture of the one-dimensional bosonic model. In the figures, a boson is located at the central red site and can hop to the neighboring site with a hopping amplitude $t$. Moreover, long-range interactions $U\gg t$ ensure that no other boson can occupy a site within a range $R_B$ of the central boson, mimicking the radius of a skyrmion. 
The boson density $n_B$ as a function of the chemical potential $\mu$ is plotted (b) for different interaction ranges $R_B$ and (c) for different hopping parameters $t$. Unless otherwise stated in the figures, the parameters are set to $t=0.25$, $U=8.0$, $R_B=2.0$.
}
\label{fig::SimpleBosonicModel}
\end{figure}

To demonstrate how a kinetic term in the Hamiltonian induced by quantum fluctuations can explain the distinct behaviors of the phase transition of the quantum skyrmion compared with the classical skyrmion, it is useful to define a simplistic real-space tight-binding model for skyrmions.
However, because skyrmions span several lattice sites, it is not possible to define a strongly localized Wannier function for a skyrmion. 
Simultaneously defining a real-space tight binding model is also not possible.
Furthermore, a skyrmionic model must adopt a real-space continuum approach, elucidating why the size of the magnetic unit cell is smaller than that of the lattice unit cell (see Fig.~\ref{fig::Skyrmion_Band_Structure}(a)).
Here, we assume that an effective tight-binding model can still describe skyrmions. 
This means that the hopping within the magnetic unit cells ($t_m$) is renormalized into hopping between the lattice sites ($t\approx (t_m)^N$). 
Consequently, in the classical limit ($N\rightarrow\infty$), the effective hopping term between two lattice sites becomes zero.
We use the following one-dimensional hard-core bosonic model,
\begin{align}
    \pazocal{H}_B &=-t\sum_{j=1}^{j=N_s-1} \left( \hat{a}^\dagger_j \hat{a}_{j+1} 
    +\hat{a}^\dagger_{j+1} \hat{a}_j
    \right)
    \nonumber\\
    &+ U\sum_{\substack{j=1,k=1,\\ \{j+k<N_s\}}}^{j=N_s, k=R_B} \hat{n}_j \hat{n}_{j+k} 
    +\mu \sum_j \hat{n}_j,
\end{align}
 where $t$ and $\mu$ are the nearest neighbor hopping amplitude and chemical potential respectively. 
 $N_s$ is the number of lattice sites.
The inter-bosonic interaction $U\gg t$ spans over several sites within a distance $R_B$ as shown in Fig.~\ref{fig::SimpleBosonicModel}(a).
The range $R_B$ is equivalent to the skyrmion radius and varying $R_B$, the boson density as a function of a chemical potential $\mu$ is plotted in Fig.\,\ref{fig::SimpleBosonicModel}(b).
Furthermore, the boson density as a function of chemical potential for different hopping amplitudes is also shown in Fig.\,\ref{fig::SimpleBosonicModel}(c).
Zero hopping amplitude corresponds to classical skyrmion and the corresponding phase diagram is discontinuous similar to Fig.\,\ref{fig::PhaseTransition}(b). 
On the other hand, non-zero hopping amplitude makes the phase transition continuous, similar to the phase transition of quantum skyrmions as in Fig.\,\ref{fig::PhaseTransition}(a).
Additionally, skyrmion number non-conserving terms may be present in the Hamiltonian due to the non-coplanar texture of skyrmions and DMI.
We discuss two types of number non-conserving terms in App.~\ref{Appendix::BC}.  
We observe that odd-order terms lead to a crossover rather than a phase transition, consistent with Ref.~\cite{LeonBalents}.
Finally, it's worth mentioning that although the hardcore bosonic model is not an exact representation of the skyrmion system, they share similar physical phenomena based on the hopping parameter $t$.

\begin{figure}[tb]
\centering
\includegraphics[width=0.45\textwidth]{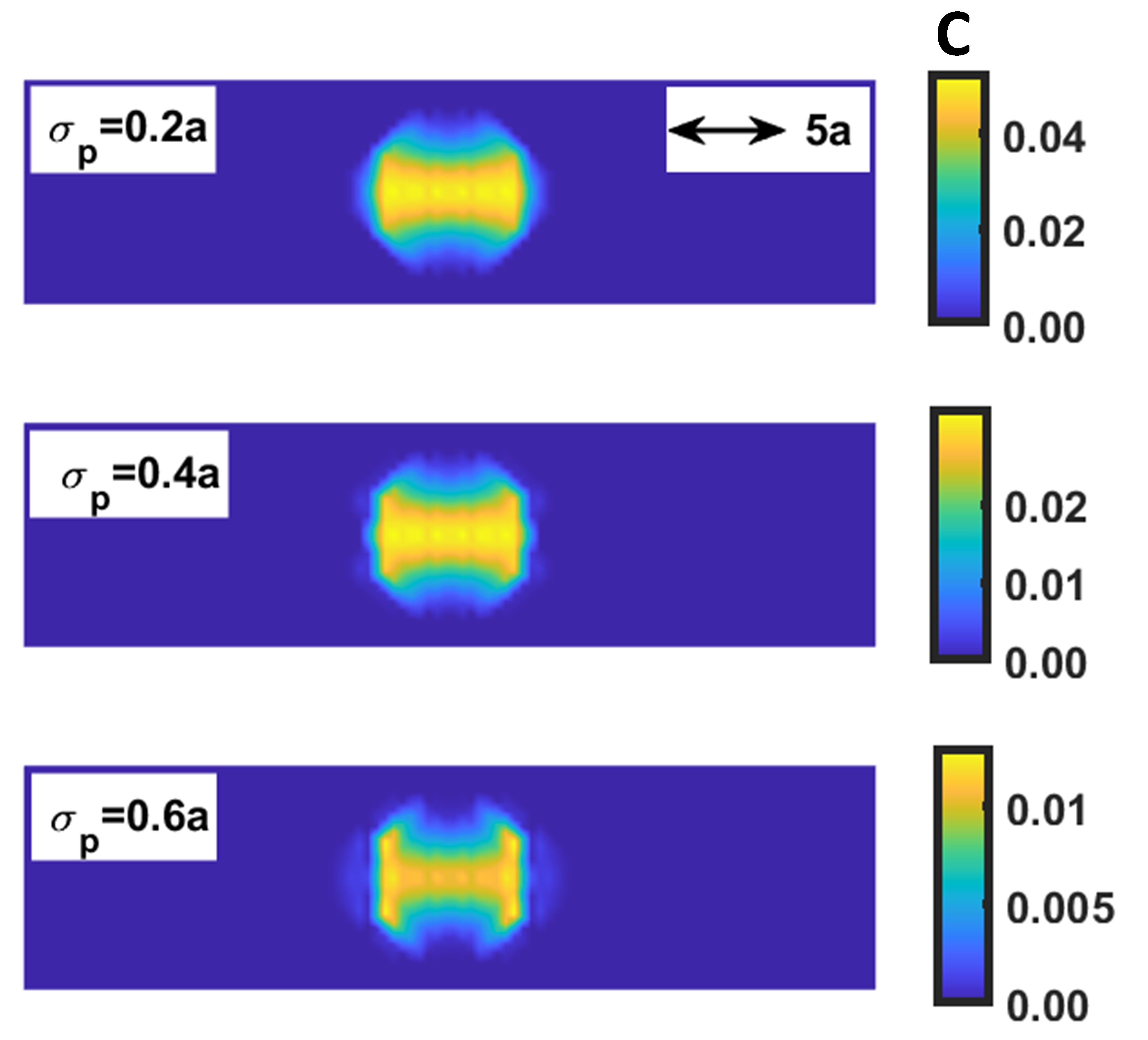} 
\caption{Entanglement (concurrence) distribution of skyrmion wave functions. As the standard deviation $\sigma_p$ of the Gaussian probability distribution is increased, the entanglement distribution becomes thinner along the horizontal line through the skyrmion center. 
}
\label{fig::EntanglementDistribution_MicroscopicModel}
\end{figure}

Next, we explain another important numerical result regarding the peculiar entanglement distribution in the SkL ground state (see Fig.~\ref{fig::EntanglementDistribution}(a)), with two main features: (1) the entanglement is highly concentrated along the horizontal line through the skyrmion center, and (2) the entanglement distribution along the circular skyrmion boundary is uneven and less concentrated near the two poles. 
Using the MPS representation of the wave function, we can calculate and plot the concurrence in Fig.~\ref{fig::EntanglementDistribution_MicroscopicModel} for different values of $\sigma_p$.
Larger values of $\sigma_p$ correspond to more superimposed single skyrmion wave functions with different centers with a Gaussian probability distribution along the horizontal direction.
The superposition along the horizontal direction results in an entanglement distribution concentrated more along the horizontal line through the skyrmion centers.
Moreover, a higher value of $\sigma_p$ leads to a thinner entanglement distribution along the horizontal line.
Furthermore, the entanglement distribution along the skyrmion boundary is also nonuniform, leaving voids at two poles.
Thus, the model wave function of the skyrmion qualitatively describes the entanglement distribution in SKL phase in Fig.~\ref{fig::EntanglementDistribution}(a).

However, the model wave function Eq.\,\ref{eq::SkyrmionicWavefunction} may not be a unique way to describe the entanglement distribution.
Some similar model wave functions can also give similar results as shown in App.~\ref{Appendix::C}.
Vivek \textit{et al.}~further show that a quantum skyrmion may contain a contribution from anti-skyrmion wave functions.
For this reason, we incorporate anti-skyrmions in the model wave function and show that the corresponding entanglement distribution, as well as real space spin polarization, do not match at all the numerical data of our DMRG simulations (see App.~\ref{Appendix::C}).
While this argument cannot rule out the presence of anti-skyrmions completely, its role in describing the numerical result is less significant, especially for the system studied in this article.

\section{\label{Section::Summary}Summary}

We've shown that skyrmionic kinetic energy mediated by quantum skyrmion-lattice interaction, results in a new skyrmionic phase SkL at zero temperature. 
The SkL phase is purely quantum mechanical and is absent in the near-zero temperature classical skyrmionic phase due to the lack of kinetic energy.
Moreover, in the quantum system, SkX and SkL phases are uniquely distinct despite being connected through a crossover rather than a real phase transition. 
The difference in skyrmion density between the SkL and SkX phases further serves as a compelling manifestation of the kinetic energy of skyrmions.
Skyrmions in the SkL phase are uncorrelated compared with the skyrmions in the SkX phase.
This is eventually reflected in both the spatial entanglement distribution and spin-spin correlation function in both the SkX and SkL phases.
In the SkL phase, the entanglement distribution is peaked around the skyrmion center, while the spin-spin correlation decays exponentially.
Whereas, the features of the SkX phase are distinctly opposite, with entanglement concentration around the skyrmion circumference and almost no decay in spin-spin correlation.

Moreover, we have utilized collective coordinate formalism and trial wavefunctions to validate the explanations for the numerical results.
Using the collective coordinate formalism, we have calculated the skyrmion band structure.
The bandwidth unequivocally represents the kinetic energy of the skyrmion and exclusively exists in a quantum system, disappearing entirely in the classical limit.
In the SkL phase, the concentrated entanglement distribution near the skyrmion center can be explained as an outcome of the quantum superposition of many skyrmionic states.

Finally, we believe it's crucial to address the potential realization of BEC of skyrmions as a future work.
In this work, we demonstrate that quantum skyrmions possess kinetic energy, conclusively paving the way toward the realization of a BEC of skyrmions, which are bosonic\,\cite{haller2024quantum}.
However, the quasi-one-dimensional system of skyrmions that we study here, does not exhibit any signature of BEC of skyrmions.
The reason behind this can be understood by comparing the skyrmions with hardcore bosons with nearest-neighbor hopping introduced in Sec.\,\ref{Section::MicroscopicDescription}.
To achieve a uniform wavefunction of hardcore bosons essential for BEC, long-range hopping is necessary such that a skyrmion is capable of skipping the closest skyrmion in a quasi-one-dimensional system.
Due to the absence of long-range hopping in the system we study, the BEC of skyrmion is not present.
So, we propose that a system with higher dimensions or quasi-one-dimensional systems with shifted boundary conditions could be a promising playground for achieving BEC of skyrmions.

Despite our inability to form a Bose-Einstein condensate (BEC) of skyrmions, our study remains important for experimentally relevant measures to quantify the quantumness of skyrmions.
The quantum phase transition from SkX to FP phase is continuous whereas in the classical case, it is a first-order in nature.
Thus, near-zero temperature experimental measurement of the magnetization vs. magnetic field curve provides valuable information about the quantum nature of skyrmions.
Furthermore, there are passive methods available for measuring entanglement in a condensed matter system\,\cite{EntanglemntWitnessInCondensedMatter}.
If in the future, an experimental technique can be developed to measure the spatially local entanglement in a system, then one can characterize the SkL phase based on its entanglement concentration near the skyrmion center.
Finally, inelastic neutron scattering experiments can differentiate between a conventional SkX phase and the quantum SkL phase by measuring the spin-spin correlation in the spin system.


\acknowledgements

DB acknowledges the use of the computational resources at the High-Performance Computing Centre (HPCC) at NTU, Singapore. AH and TLS acknowledge financial support from the National Research Fund Luxembourg under grants C22/MS/17415246/DeQuSky and C20/MS/14764976/TOPREL.

\pagebreak
\appendix

\section{\label{Appendix::A} Quantum Simulation For Higher Spins}

\begin{figure}[tb]
\centering
\includegraphics[width=0.45\textwidth]{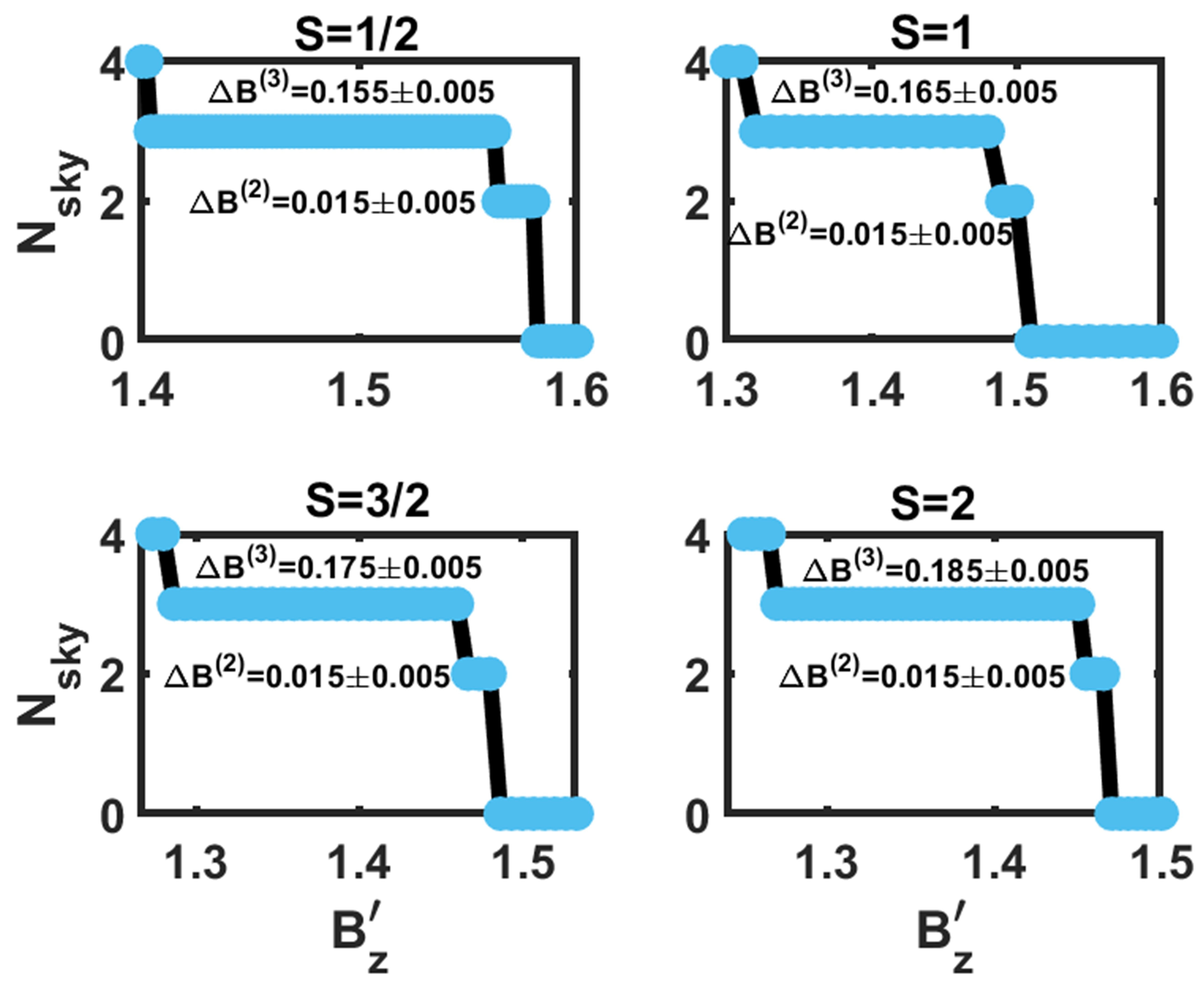} 
\caption{
The number of skyrmions as a function of normalized magnetic field $B_z^\prime=B_z/S$ for different spin values for system size $L^x=27$, $L^y=9$.
}
\label{fig::QuantumPhaseTransition_DifferentSpins}
\end{figure}

In the main text, we claimed that the comparison between the results of quantum simulation of the same model Hamiltonian for different spin is not a straightforward comparison of quantum nature of skyrmion.
Because, increase in spin ($S\rightarrow\infty$, classical limit) of a system, decreases the radius of the skyrmion ($R\rightarrow 0$, quantum limit).
This is illustrated in this appendix.

Using the model Hamiltonian Eq.\,\ref{eq::System_Hamiltonian}, we plot the number of skyrmions as a function of normalized magnetic field $B_z^\prime=B_z/S$ in Fig.\,\ref{fig::QuantumPhaseTransition_DifferentSpins}.
 We use a small system size $L^x=27$ because of numerical complexity increases with the spin value $S$, which leads to discrete looking phase transition compared with the phase transition in Fig.\,\ref{fig::PhaseTransition} in the main text.
However, by comparing the sizes of the plateaus, we can still comment on the stability of SkL phase.
In figure Fig.\,\ref{fig::QuantumPhaseTransition_DifferentSpins}, we observe that the size of the plateau increases with the spin of the system, indicating increase in the stability of SkL phase. 
It demonstrates that the SkL phase becoming more robust towards the classical $S\gg 1/2$ limit, which is counter-intuitive.

\begin{figure}[tb]
\centering
\includegraphics[width=0.45\textwidth]{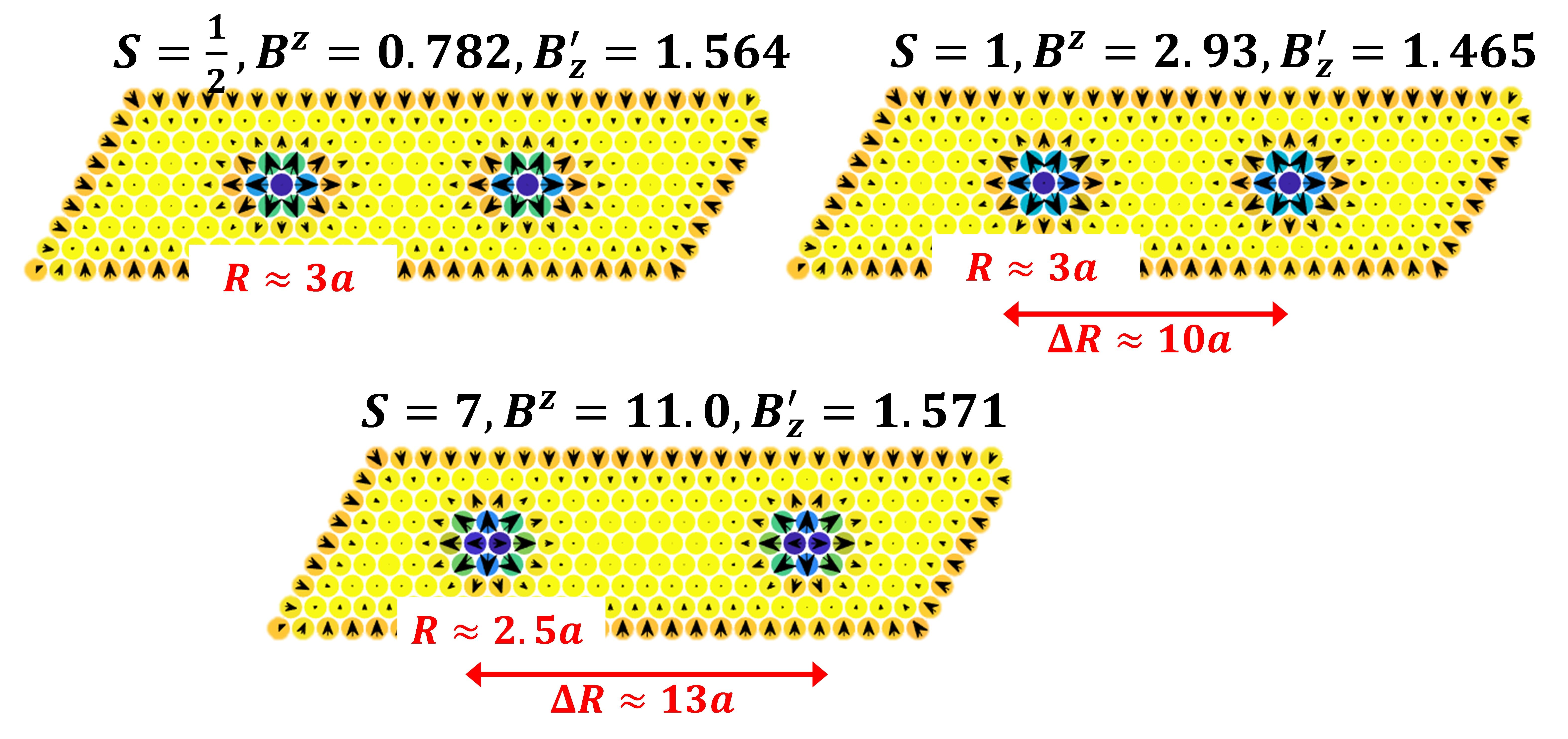} 
\caption{
Size of skyrmion for different spin values.
}
\label{fig::RadiusOfSkyrmions}
\end{figure}

However, the radius of the skyrmion is also an important parameter to determine the quantum nature of a skyrmion, which needs to be considered.
In Fig.\,\ref{fig::RadiusOfSkyrmions}, we show that the radius of a skyrmion decreases with in increase in spin value $S$ of a system , as higher magnetic field is required to stabilize the skyrmionic phase.
Decrease in radius does not only increase the quantum nature of a skyrmion, but also decrease the skyrmion-skyrmion interaction, which leads to stable of SkL phase.

\section{\label{Appendix::B} Skyrmion-Skyrmion Interaction}

\begin{figure}[tb]
\centering
\includegraphics[width=0.45\textwidth]{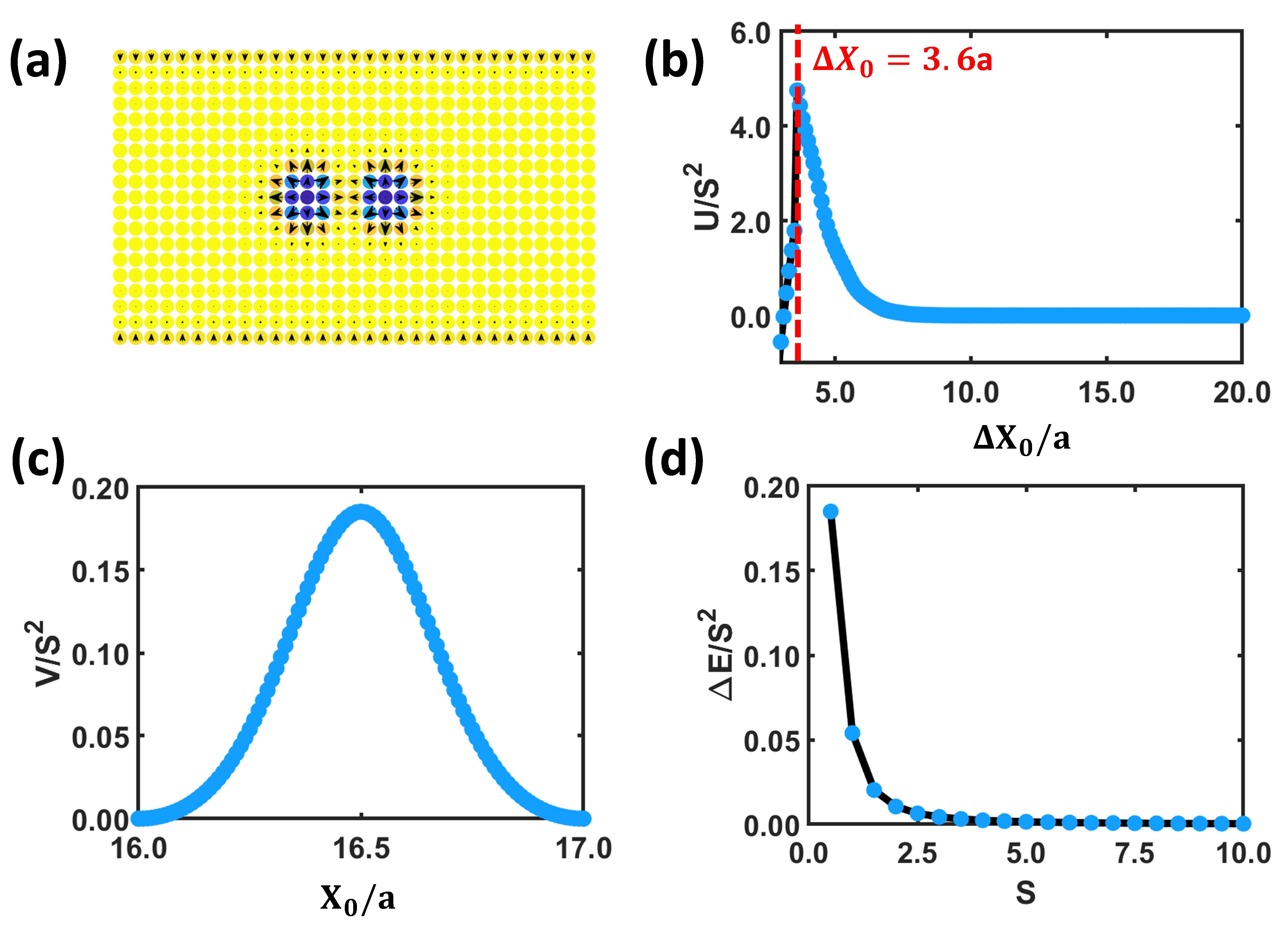} 
\caption{(a) Two interacting Skyrmions with distorted skyrmion boundary and domain.  
(b) Skyrmion-skyrmion interaction U as a function of distance between skyrmion centers $\Delta X_0$. The data interaction for $\Delta X_0<3.6a$ is redundant as the Gaussian pinning magnetic field is unable to produce two distinct skyrmions in that regime.
(c) Potential profile for a skyrmion produced by Gaussian pinning magnetic field. 
(d) Bandwidth of lowest skyrmion band as a function of spin $S$.
The parameters for pinning magnetic field $(B_0^z,\, \sigma_B)$ is fixed to $(2.0,\,1.0)$ for all the plots. 
}
\label{fig::Skyrmion_Interaction}
\end{figure}

In this appendix, we study the nature of skyrmion-skyrmion interaction. 
To simulate skyrmion-skyrmion interaction one needs to consider the interaction driven deformation of skyrmion boundary and the domain in between two skyrmions (see Fig.\,\ref{fig::Skyrmion_Interaction}(a)), which is hard to describe by an analytical expression similar to Eq.\,\ref{eq::Classical_Skyrmion_Config}.
Thus, we use the system described by the Hamiltonian Eq.\,\ref{eq::System_Hamiltonian} in the FP regime $B_z=0.8D$ along with an extra Gaussian pinning potential,
\begin{equation}
    B_{\text{pin}}^z(x,y)=
    \sum_{i}
    B_0^z\exp\left(-\frac{\left(x-X_0^i\right)^2+(y-Y_0)^2}{2(\sigma_B)^2}\right),
    \label{eq::Pinning_Potential}
\end{equation}
which pins two skyrmions at the centers $(X_0^1,\,Y_0)$ and $(X_0^2,\,Y_0)$ for $i\in(1,2)$ and $Y_0=(L^y+1)/2$.
By varying the position of one skyrmion while fixing the other, the interaction $U$ is calculated as a function of distance ($\Delta X_0=|X_0^1-X_0^2|$) between the two skyrmions,
\begin{equation}
    U(\Delta X_0)=V(\Delta X_0)-V(X_0^1)-V(X_0^2),
\end{equation}
where $V(\Delta X_0)$ and $V(X_0^i)$ represent the energy of the system with two and one skyrmions, respectively, with respect to the energy of the system without a skyrmion.
The plot Fig.\,\ref{fig::Skyrmion_Interaction}(b) shows the interaction strength of the skyrmion as distance between the two skyrmions increases.
In conclusion, the skyrmion interaction is short ranged and decreases rapidly with distance.

Furthermore, in order to enumerate the skyrmion band, we use the pinning potential Eq.\,\ref{eq::Pinning_Potential} with $i\in(1)$ to pin one skyrmion at position $(X_0^1=X_0, Y_0)$ and evaluate the potential profile $V(X_0)$ as shown in Fig.\,\ref{fig::Skyrmion_Interaction}(c).
Using the potential profile, the band width of the lowest skyrmion band is calculated as a function of spin (see Fig.\,\ref{fig::Skyrmion_Interaction}(d)) which shows a rapid decrease in band width with the spin of the system.
Although, in the Fig.\,\ref{fig::Skyrmion_Interaction}(b), it seems that the skyrmion interaction is short ranged, the longer distance interaction becomes very important for high spin values as the skyrmion kinetic energy also decreases rapidly with spin. 
Therefore, for large spin values, a SkL phase with higher skyrmion number has lower stability which explains the numerical results Fig.
Moreover, the SkL phase only appears within a short range of magnetic field for higher spins for the same reason.


\section{\label{Appendix::BC} Effect of additional number non-conserving terms in the 1D bosonic Hamiltonian}

\begin{figure}[tb]
\centering
\includegraphics[width=0.4\textwidth]{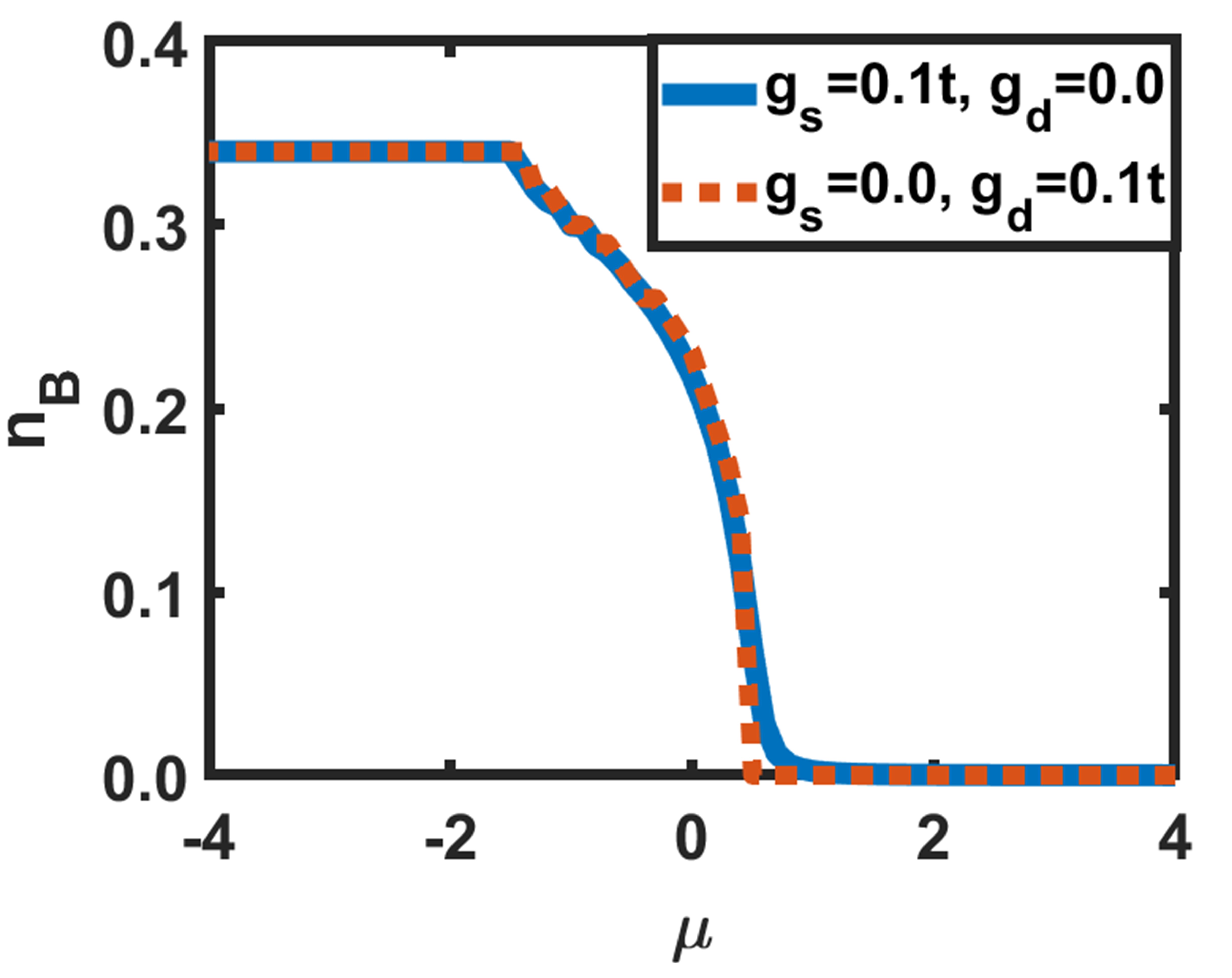} 
\caption{The density of boson as a function of chemical potential with single (or pair) creation-annihilation operators shown in blue (or red) color.
}
\label{fig::SimpleBosonicModel_Appendix}
\end{figure}

 In the main text, we defined a simple one-dimensional hard-core bosonic model to describe the distinction between continuous and discrete phase transition of quantum and classical skyrmions respectively.
 However, skyrmion number non-conserving terms may be present in the effective skyrmionic Hamiltonian and to take that into account we added single and pair creation-annihilation terms in the Hamiltonian,
 \begin{align}
      \pazocal{H}_B &=
      g_s \sum_j \left(
      \hat{a}^\dagger_j+\hat{a}
      \right)
      -t\sum_j \left(
      \hat{a}^\dagger_j \hat{a}_{j+1}
      +\hat{a}^\dagger_{j+1} \hat{a}_j
      \right)
      \nonumber\\
      &+g_d \sum_j \left( 
      \hat{a}^\dagger_j \hat{a}_{j+1}^\dagger
      +\hat{a}_j \hat{a}_{j+1}
      \right)
      \nonumber \\
    &+ U\sum_{j,k=1}^{k=R_B} \hat{n}_j \hat{n}_{j+k} 
    +\mu \sum_j \hat{n}_j.
 \end{align}
The boson density as a function of chemical potential is shown in Fig.\,\ref{fig::SimpleBosonicModel_Appendix}.
It is noticeable that near the critical point, the order parameter for $g_s=0.0$ suddenly drops to zero, while the order parameter for $g_d=0.0$ asymptotically approaches zero.
Thus, It is evident that the presence of non-conserving terms with odd order leads to a crossover instead of a proper phase transition\,\cite{LeonBalents}.

\section{\label{Appendix::C} Entanglement Distribution}

\onecolumngrid

\begin{figure}[tb]
\centering
\includegraphics[width=0.95\textwidth]{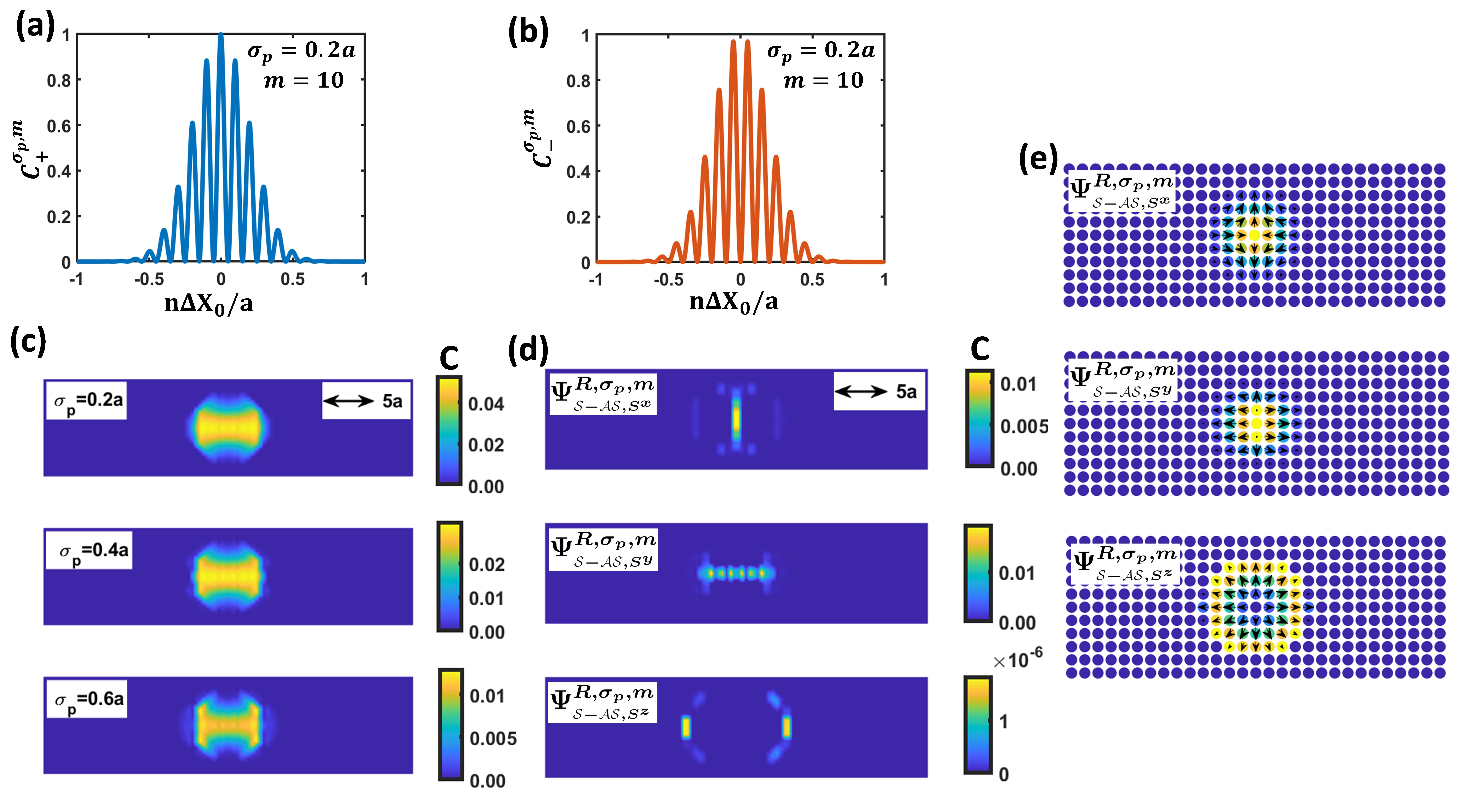} 
\caption{
The coefficients $C_+^{\sigma_p, m}$ and $C_-^{\sigma_p, m}$, which represent skyrmion and antiskyrmion probability distributions, are plotted as functions of space in (a) and (b) respectively.
(c) Entanglment distribution for skyrmionic wavefunction $\Psi_{{}_\pazocal{S}}^{R,\sigma_p, m}$ for different $\sigma_p$ values. (d) Entanglment distribution and real space spin polarization for skyrmionic-antiskyrmionic wavefunction $\Psi_{{}_{\pazocal{S}-\pazocal{AS}, S^\alpha} }^{R,\sigma_p,m}$ (where $\alpha\in (x,y,z)$) are shown in (d) and (e) respectively.
Unless otherwise stated, all figures use $R=4.0a$, $\sigma_p=0.5a$, $m=10$, $\Delta X_0=0.001$.
}
\label{fig::Entanglement_Distribution_Appendix}
\end{figure}
\twocolumngrid

In the main text, we show the entanglement distribution for a specific model wavefunction described by Eq.\,\ref{eq::SkyrmionicWavefunction}. 
In this apppendix, we use the following model wave function,
\begin{align}
    \Psi_{{}_\pazocal{S}}^{R,\sigma_p,m}(X_0)&=\frac{1}{\sqrt{N}}
    \sum_{n\in \mathbb{Z}} 
    C^{\sigma_p, m}_+\left(n\Delta X_0\right)
    \Psi^R_{{}_\pazocal{S}}(X_0+n\Delta X_0),
    \label{eq::SkyrmionicWavefunction_Appendix}
\end{align}
where, the coefficient,
\begin{align}
    C_+^{\sigma_p, m}\left(n\Delta X_0\right)=\frac{1}{2} 
    &\exp(-\frac{1}{2}\left(\frac{n\Delta X_0}{\sigma_p}\right)^2)
    \nonumber\\
     &\times(1+\cos(2\pi mn\Delta X_0)).
\end{align}
The spatial profile of the coefficient $C_+^{\sigma_p, m}$ is depicted in Fig.\,\ref{fig::Entanglement_Distribution_Appendix}(a), which shows that the oscillatory nature due to cosine term and exponential decay due to Gaussian term both are present.
Furthermore, the entanglement distribution of the wavefunction Eq.\,\ref{eq::SkyrmionicWavefunction_Appendix} is plotted in Fig.\,\ref{fig::Entanglement_Distribution_Appendix}(c) by varying parameters $\sigma_p$.
The entanglement profile matches exactly with the profile in Fig.\,\ref{fig::EntanglementDistribution_MicroscopicModel}.
Thus oscillatory nature of the probability distribution does not affect the nature of entanglement distribution.

Vivek \textit{et al.}~show that, the wavefunction for a skyrmion might be a superposition of skyrmion and anti-skyrmion.
We incorporate antiskyrmions in the following manner,
\begin{align}
    \Psi_{{}_{\pazocal{S}-\pazocal{AS}, S^\alpha} }^{R,\sigma_p,m}(X_0)=\frac{1}{\sqrt{N}}
    \sum_{n\in \mathbb{Z}} 
    \left[ C^{\sigma_p, m}_+\left(n\Delta X_0\right)
    \Psi^R_{{}_\pazocal{S}}(X_0+n\Delta X_0)
    \right.
    \nonumber\\
    +\left.
    C^{\sigma_p, m}_-\left(n\Delta X_0\right)
    \Psi^R_{{}_{\pazocal{AS},S^\alpha}}(X_0+n\Delta X_0)
    \right],
    \label{eq::SkyrmionicAntiSkyrmionicWavefunction_Appendix}
\end{align}
where, the coefficient,
\begin{align}
    C_-^{\sigma_p, m}\left(n\Delta X_0\right)=\frac{1}{2} 
    &\exp(-\frac{1}{2}\left(\frac{n\Delta X_0}{\sigma_p}\right)^2)
    \nonumber\\
     &\times(1-\cos(2\pi mn\Delta X_0)),
\end{align}
is converse of $C_+^{\sigma_p, m}$ and shown in fig.\,\ref{fig::Entanglement_Distribution_Appendix}(b). Furthermore, the antiskyrmionic wavefunction is derived from the skyrmionic wavefunction by inverting one of the spin component of the skyrmion,
\begin{equation}
    \Psi^R_{{}_\pazocal{S}} 
    \xrightarrow{S^\alpha\rightarrow -S^\alpha}
    \Psi^R_{{}_{\pazocal{AS},S^\alpha}}.
\end{equation}
Thus there are three possible definitions of the model wavefunction Eq.\,\ref{eq::SkyrmionicAntiSkyrmionicWavefunction_Appendix} by inverting one of the three spin components.
The entanglement distribution as well as the real space spin polarization of the wavefunction Eq.\,\ref{eq::SkyrmionicAntiSkyrmionicWavefunction_Appendix} is shown in Fig.\,\ref{fig::Entanglement_Distribution_Appendix}(d) and (e) respectively.
Although, the entanglement distribution for wavefunctions
$\Psi_{{}_{\pazocal{S}-\pazocal{AS}, S^x} }^{R,\sigma_p,m}$
and 
$\Psi_{{}_{\pazocal{S}-\pazocal{AS}, S^y} }^{R,\sigma_p,m}$
may have a close relevance with the numerical results for SkL phase in Fig.\,\ref{fig::EntanglementDistribution}(a), 
the real space spin polarization is completely different compared with Fig.\,\ref{fig::PhaseTransition}(c).
Therefore, we conclude that the presence of antiskyrmion for our Hamiltonian system is not very significant.

\bibliographystyle{apsrev4-PRX}

%

\end{document}